\documentclass[10pt,onecolumn,superscriptaddress]{article}

\usepackage{graphics}
\usepackage{graphicx}
\usepackage{subfig}
\usepackage{color}
\usepackage[english]{babel}
\usepackage{amsmath}
\usepackage{authblk}
\usepackage{epstopdf}

\begin{document}

\title{Effects of memristor-based coupling in the ensemble of FitzHugh-Nagumo elements}
\author[1]{Alexander G. Korotkov}
\author[1]{Tatiana A. Levanova}
\author[2]{Alexey O. Kazakov}
\affil[1]{Control Theory Department, Institute of Information Technologies, Mathematics and Mechanics, Lobachevsky University, Gagarin ave. 23, Nizhny Novgorod, 603950, Russia}
\affil[2]{Faculty of Informatics, Mathematics, and Computer Science, National Research University Higher School of Economics, Bolshaja Pecherskaja Str. 5/12, Nizhny Novgorod, 603155, Russia}
\date{\today}

\maketitle

\begin{abstract}
In this paper, we study the impact of electrical and memristor-based couplings on some neuron-like spiking regimes, previously observed in the ensemble of two identical FitzHugh-Nagumo elements with chemical excitatory coupling. We demonstrate how increasing strength of these couplings affects on such stable periodic regimes as spiking in-phase, anti-phase and sequential activity. We show that the presence of electrical and memristor-based coupling does not essentially affect regimes of in-phase activity. Such regimes do not changes remaining stable ones. However, it is not the case for  regimes of anti-phase and sequential activity. All such regimes can transform into periodic or chaotic ones which are very similar to the regimes of in-phase activity. Concerning the regimes of sequential activity, this transformation depends continuously on the coupling parameters, whereas some anti-phase regimes can disappear via a saddle-node bifurcation and nearby orbits tend to regimes of in-phase activity.
Also, we show that new interesting neuron-like phenomena can appear from the regimes of sequential activity when increasing the strength of electrical and/or memristor-based coupling. The corresponding regimes can be associated with the appearance of spiral attractors containing a saddle-focus equilibrium with homoclinic orbit and, thus, they correspond to chaotic motions near the invariant manifold of synchronization, which contains all in-phase limit cycles. Such new regimes can lead to the emergence of extreme events in the system of coupled neurons. In particular, the interspike intervals can become arbitrarily large when orbits (corresponding to this regime) pass very close to the saddle-focus. Finally, we show that the further increase in the strength of electrical coupling and/or memristor-based coupling leads to decreasing interspike intervals and, thus, it helps to avoid such extreme behavior.
\end{abstract}

\section{Introduction}
\label{intro}

The neural system in humans and animals consists of a large number of neurons organized into hierarchical structures and capable of reproducing complex collective behaviour, which is a specific response of patterns of electrical activity of neural cells \cite{Nicholls2011}. Such electrical activity can appear in various forms, such as subthreshold impulses, tonic spikes, bursts, and even chaotic oscillations as a reaction to the simulation of various neurons by current, and it can also propagate through various types of connections.

Chemical synaptic couplings are the most common type of couplings in the nervous system. They are always unidirectional, that is, the signal is transmitted from the presynaptic element to the postsynaptic one. The strength of the postsynaptic element's response depends on the amplitude of the membrane potential in the presynaptic element. If this amplitude is less than a certain threshold value, then postsynaptic element demonstrates no significant response. If the amplitude of the membrane potential is greater than a certain threshold value, then, depending on the type of coupling (excitatory or inhibitory), the postsynaptic element becomes excited or suppressed \cite{pereda}. Note that such couplings are inertial, i.e. they have specific times of action. Neuronal ensembles with chemical synapses can demonstrate various types of complex activity, including synchronous and cluster regimes \cite{sync}, sequential activity \cite{Afraimovich2004,Levanova2013} and other non-trivial regimes \cite{chimera}.

It should be noted that different types of synchronization play crucial role in the activity of the brain and nervous system. Synchronization in some areas and clusters of neurons in the brain \cite{20} is associated both with various complex biological functions (such as memory \cite{94} and mental alertness \cite{21}) and various pathological processes \cite{epilepsy,parkinson}. Many important studies have been devoted to this important phenomenon, for example, in papers \cite{32}-\cite{34} the authors study the synchronization in different realistic neural systems consisting of a large number of non-identical elements.

Neural ensembles consist of a large number of elements, and the complete synchronization in an ensemble containing only chemical synaptic couplings can be difficult to implement due to the diversity of types of neuron  \cite{nosync}. Consequently, other possible efficient and simple methods for transmitting signals between neurons should be studied in more details.

One of the quite appropriate options for such couplings are those based on gap junctions in electrical synapses. Signal transmissions in electrical synapses, unlike chemical ones, are usually bidirectional \cite{151,223}. Furthermore, the neuron can generate a response not only to the action potential, but also to the subliminal activities, such as the synaptic potentials \cite{224} and spontaneous oscillations \cite{210}. The additionally introduced electrical synapses have a huge impact on the dynamics of the ensemble. They are of extreme plasticity and able to change their coupling strength under different physiological conditions \cite{229}-\cite{233}. In their  turn, these changes can quickly reconfigure the full network, for example, as it can be observed in processes happened in the retina \cite{187}. A number of studies have shown that electrical synapses can efficiently synchronize the network activity \cite{234}-\cite{Belykh2017}, which ultimately helps to model complex nervous activities, such as the behaviour of individum, performing of the cognitive tasks, and the consciousness \cite{239}-\cite{241}.

Note  that it is necessary to take into account such global effects as the electromagnetic induction effect \cite{MaTang2017} in the studies of complex collective behaviour of neurons. In the papers \cite{LvWang2016,LvMa2016} it was shown that time-dependent changes in the intracellular and extracellular concentration of ions can cause electromagnetic induction, and this effect can be described using magnetic flux in accordance with the law of electromagnetic induction. The current exchange in the  ion channels can also change the distribution of the electromagnetic field in the environment. The current induced by electromagnetic induction can modulate the membrane potential of a neuron. A number of studies use the so-called connection through a common electromagnetic field, simulated using memristor-based synapses \cite{LvWang2016,Chua1971}, to describe this phenomenon. This way of modelling the couplings makes it possible to mathematically describe the memory effect that exists in real neural networks, since the conductivity of the memristor depends on some features of the external signal received by the element \cite{13}-\cite{16}. It was also shown that such electromagnetic fields can cause various regimes of electrical activity \cite{LvMa2016} and phase synchronization \cite{37}, even in the presence of noise. Despite the exceptional relevance of this topic, there are relatively few works devoted to the study of such couplings and their influence on the dynamics of neural ensembles. Among these works, first of all, we note the work \cite{Volos2015}, in which, using Hindmarsh-Rose and FitzHugh-Nagumo models as an example, it was shown that many interesting dynamical effects related to the behaviour of neurons can be described using the memristor as an electrical synapse. In \cite{MaMi2017} the synchronization in a small ensemble of two connected Hindmarsh-Rose neurons was studied taking into account electromagnetic induction and magnetic flux. It was shown that, under certain conditions, phase synchronization for chaotic time series of the membrane potential can be observed in the system.

The aim of this work is to study the impact of electrical and memristor-based couplings (connections through a common field) on the dynamics of a minimal ensemble of neuron-like systems with chemical (synaptic) excitatory couplings. Previously, the authors proposed in \cite{KorotkovCNSNS2019} a new method for modelling chemical synaptic couplings using a rectangular function, taking into account the strength of the coupling, as well as the time of the beginning of the impact of the connection and the duration of this impact. Although this model is quite simple from a computational point of view, it allows, nevertheless, to organize such a phenomenological modelling the actions of chemical synapses, which is in good agreement with the biological principles of their functioning \cite{destexhe1994efficient}. This model was tested in the studies \cite{KorotkovIFAC2018,KorotkovCNSNS2019}, where the FitzHugh-Nagumo system ensemble was studied only with chemical (synaptic) excitatory connections. In these papers, a number of coexisting regimes of different types were also found in the model depending on the relations between the values of coupling parameters: regular spike in-phase, anti-phase and sequential activity regimes with different sequence of activation of elements, as well as chaotic regimes. In the present paper, we propose an extended and even more accurate model of an ensemble of coupled neurons and examine how the additional couplings (electrical couplings and couplings via the common field) allow us to more effectively manage the dynamics of the ensemble. To confirm this, we carry out one-parameter bifurcation analysis of the model in order to reveal the nature of the effect of additional couplings on previously detected regimes of activity. Using this toolbox, we study how regimes of in-phase, anti-phase and sequential activity can transform into other regimes of neuron-like activity in the presence of additional couplings. 

In this paper we also show that the presence of electrical and/or memristor-based coupling can lead to the emergence of extreme events associated with the appearance of the spiral attractors containing a saddle-focus equilibrium with its homoclinic orbit. In this case interspike intervals can become arbitrarily large when orbits (corresponding to this regime) pass near the saddle-focus.

\section{The model}
\label{sec:model}

In order to study and describe effects that emerge due to adding electrical synapses and memristor-based couplings to the neuronal ensemble with chemical synaptic couplings we need to build the correct mathematical model. This model should be, on the one hand, biologically relevant one, and on the other hand, not very complicated. In this section we introduce such a model. This model can be considered as an extension of the model proposed in \cite{KorotkovCNSNS2019} which describes interaction of two identical neurons by means of excitatory chemical coupling. First, let us introduce this basic systems and recall some important details concerning neuron-like regimes in it. Each element in the ensemble is described by the FitzHugh-Nagumo system \cite{FHN}. Thus, the system of two coupled elements takes the following form
\begin{equation} \label{ensemble_chemical}
\begin{cases}
\epsilon \mathop{x_1}\limits^\cdot = x_1 - {x_1}^3/3 - y_1 + I(\phi_2)  \\
\mathop{y_1}\limits^\cdot = x_1 - a \\
\epsilon \mathop{x_2}\limits^\cdot = x_2 - {x_2}^3/3 - y_2 + I(\phi_1) \\
\mathop{y_2}\limits^\cdot = x_2 - a \\
\end{cases}.
\end{equation}
Here $x_i$ ($i = 1,2$)  are one-dimensional variables, which describe the dynamics of membrane potential of $i$-th element, $y_i$ is called the recovery variables, which set slow negative feedback for $i$-th element, also $\epsilon$ is a small parameter, $0 < \epsilon << 1$. In further studies we assume that each of coupled elements is initially (before we set all couplings) in an excitable regime ($a = -1.01$). We also fix $\epsilon = 0.01$. 

Chemical synaptic couplings are given by the following formula
\begin{equation} \label{chemical_coupling}
I(\phi) = \frac{g}{1 + e^{k(\cos (\delta/2) - \cos (\phi - \alpha - \delta/2))}}
\end{equation}
where $\phi=\arctan\frac{y}{x}$ is measured in degrees, $0 \leq \phi < 360^\circ$, the parameter $g$ describes the strength of chemical synaptic coupling between elements. For suitable sufficiently large values of $k$, the coupling function $I(\phi)$ is a smooth function that approximates very well rectangular wave-pulses. Some additional details, as well as justification of applicability of function \eqref{chemical_coupling} for modelling chemical synaptic coupling can be found in \cite{KorotkovCNSNS2019}. In further research we will take the following values of chemical coupling parameters: $k = 50$, $g = 0.1$, $\delta = 50^\circ$.

\begin{figure}[h!]
	\centering
	\subfloat[$L_{anti}$: $\alpha = 210^\circ$, $\delta = 50^\circ$]
	{
		\includegraphics[width = .5\linewidth]{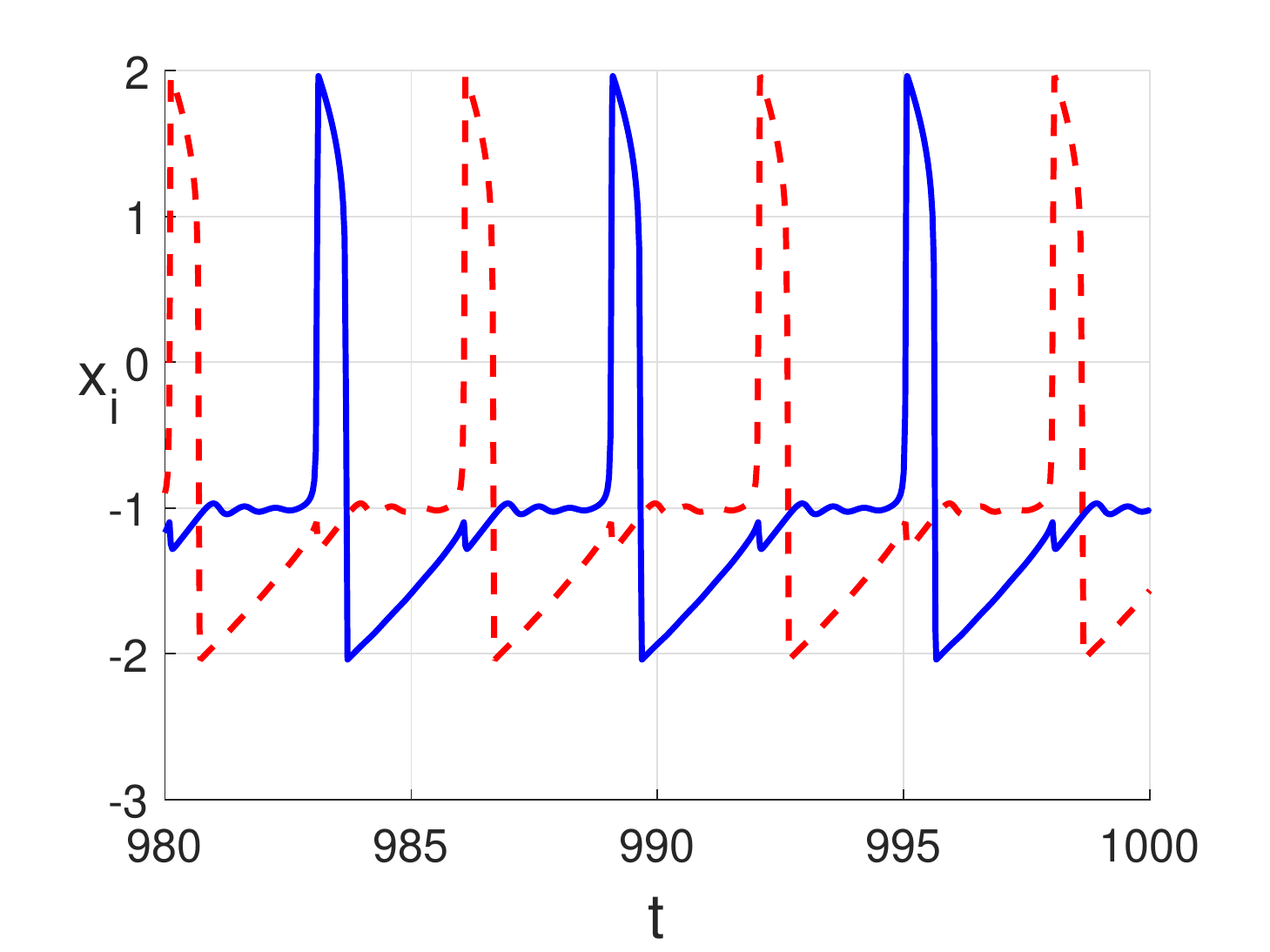}
		\label{time_serieses_a}
	}
	\subfloat[$L_{in}$: $\alpha = 210^\circ$, $\delta = 50^\circ$]
	{
		\includegraphics[width = .5\linewidth]{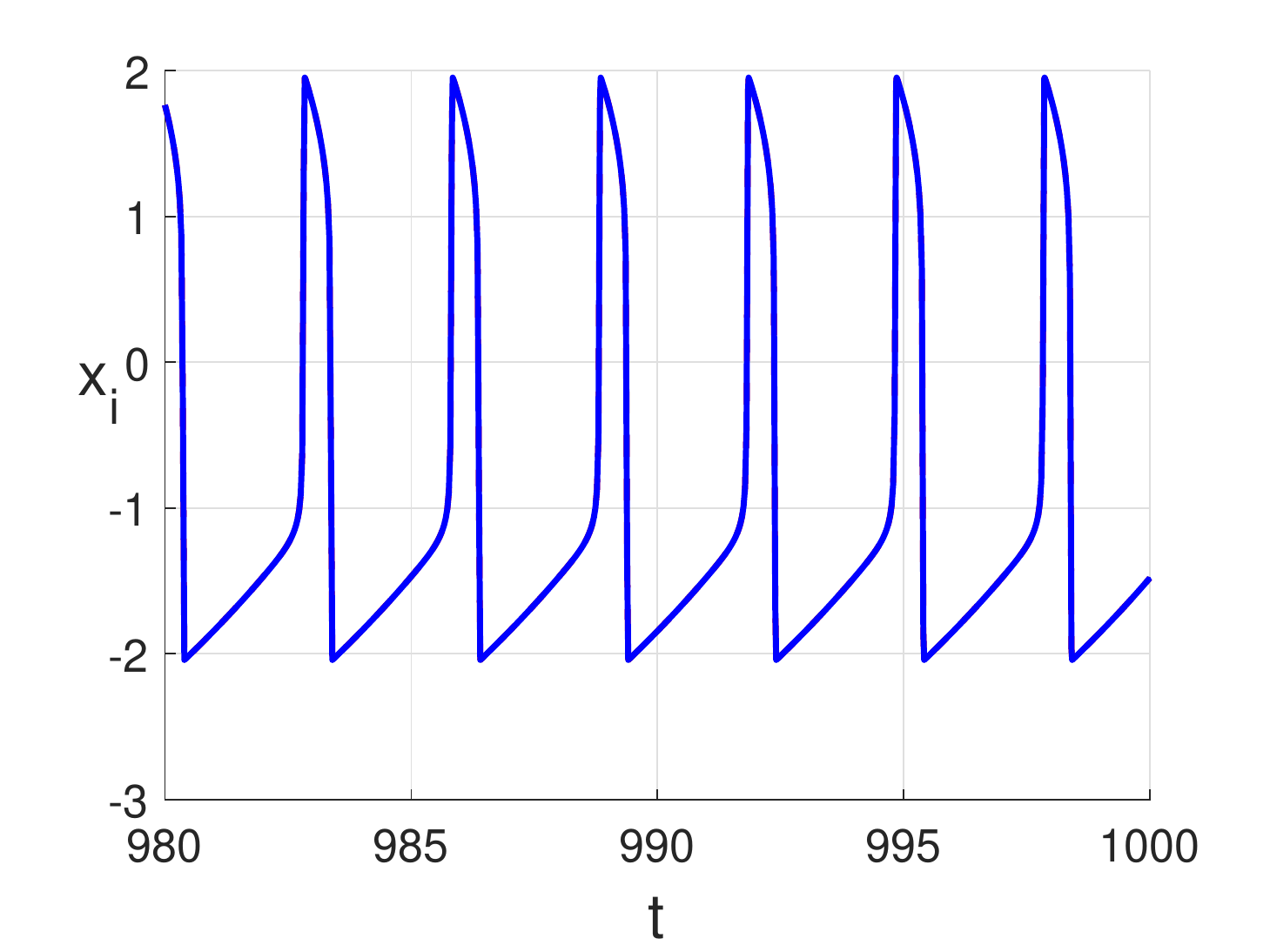}
		\label{time_serieses_b}
	}\\
	\subfloat[$L_{12}$: $\alpha = 157^\circ$, $\delta = 50^\circ$]
	{
		\includegraphics[width = .5\linewidth]{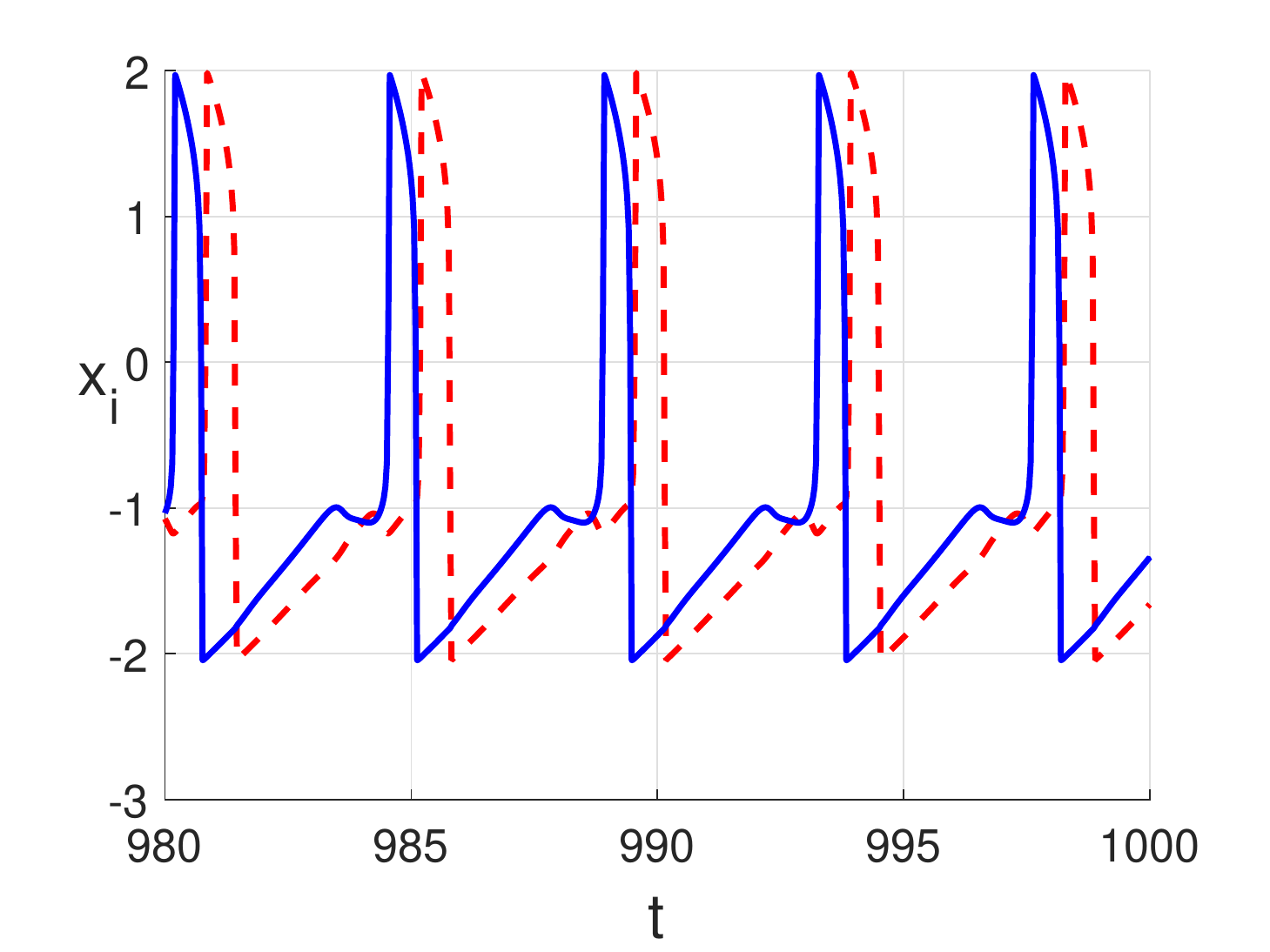}
		\label{time_serieses_c}
	}
	\subfloat[$L_{1221}$: $\alpha = 157^\circ$, $\delta = 50^\circ$]
	{
		\includegraphics[width = .5\linewidth]{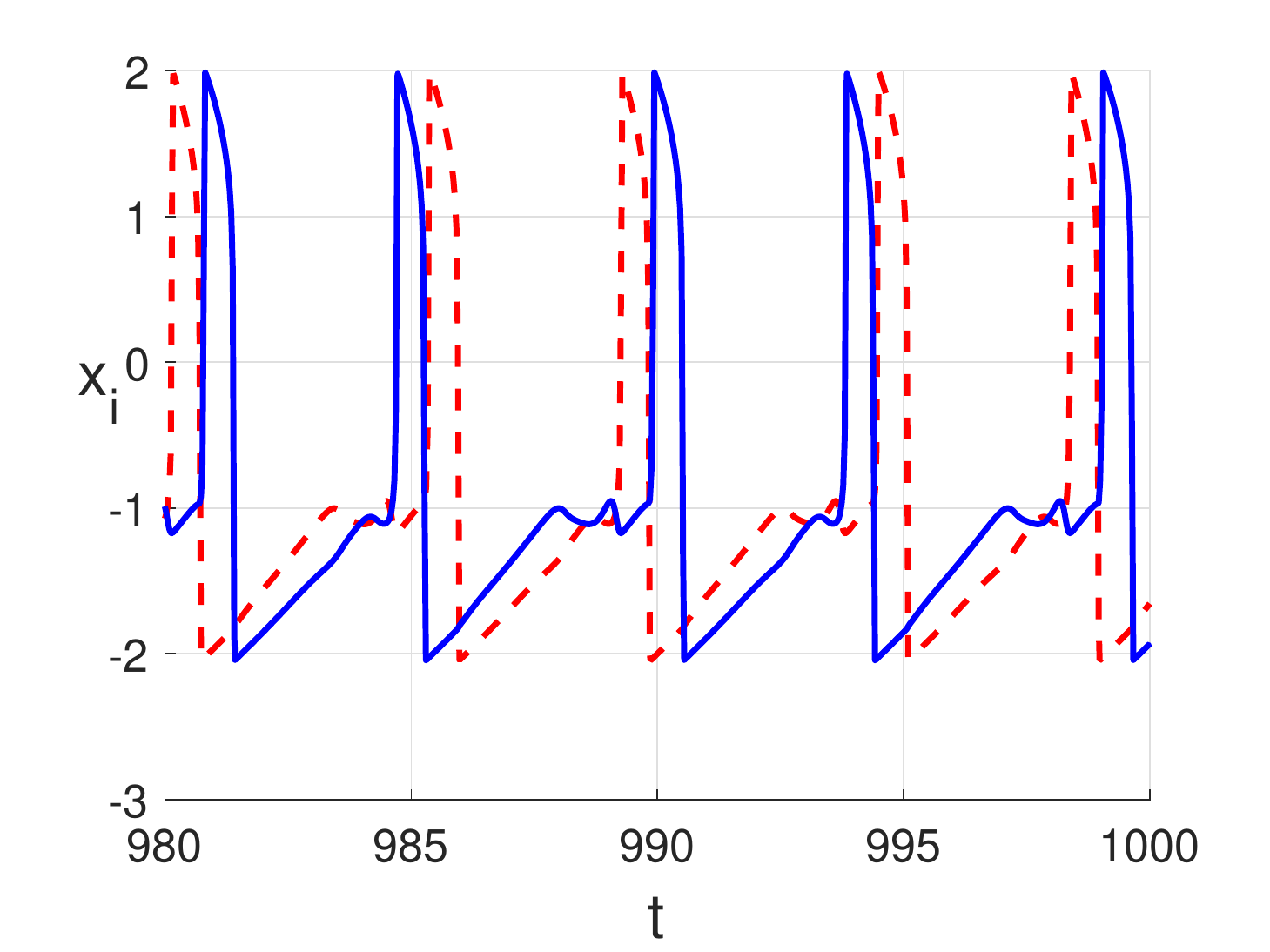}
		\label{time_serieses_d}
	}
	\caption{Time series for variables $x_1(t)$ and $x_2(t)$ for different types of neuron-like activity regimes. (a) Anti-phase tonic spiking $L_{anti}$. (b) In-phase tonic spiking $L_{in}$. (c) Sequential tonic spiking activity with first leading element $L_{12}$. (d)  Sequential tonic spiking activity with switching order of activation of elements 1-2-2-1.}
	\label{time_serieses}
\end{figure}

In \cite{KorotkovCNSNS2019} it was also shown that in ensemble \eqref{ensemble_chemical} various regimes of neuron-like activity can be generated, including regular regimes of in-phase, anti-phase and sequential activity, as well as chaotic anti-phase regimes. Here, under the in-phase regime, we mean such synchronous oscillations, in which the states of both elements coincide (see, e.g. figure~\ref{time_serieses_b}); the anti-phase regime is understood as the synchronous oscillations, where the state of the first element coincides with the state of the second element shifted to half the period: $ x_1(t) = x_2 (t \pm T/2), y_1(t) = y_2 (t \pm T/2) $, where $T$ is the period of the limit cycle (see, e.g. figure~\ref{time_serieses_a}). Sequential activity regime is also understood as a synchronous regime in which activity switches between elements, so that elements sequentially generate action potentials, and then remain suppressed for a while. Figure~\ref{time_serieses_c} shows the regime of sequential activity $L_{12}$ with the first leading element (when the spike in the first element activates the second element). In figure~\ref{time_serieses_d} more complicated regime of sequential activity $L_{1221}$ is presented.

The extended model in addition to the described above synaptic coupling takes into account electrical and memristor-based couplings \cite{MaTang2017}. In this case neurons can exchange signals by setting different electromagnetic field and magnetic flux coupling arise here \cite{LvWang2016,MaMi2017}. In the framework of this approach a flux-controlled memristor \cite{Chua1971} with the memductance
\begin{equation} \label{memristive_coupling}
\rho(\phi) = \frac{dq(\phi)}{d\phi} = k_1 + k_2\phi^2,
\end{equation}
depending on memristor parameters $k_1$ and $k_2$ is used to simulate such coupling \cite{Sharifi2010,Volos2015}.

Finally the model of two FitzHugh-Nagumo elements interacting via chemical, electrical and memristor-based couplings takes the following form
\begin{equation} \label{ensembles}
\begin{cases}
\epsilon \mathop{x_1}\limits^\cdot = x_1 - {x_1}^3/3 - y_1 + I(\phi_2) + \rho(z) \cdot(x_2 - x_1) \\
\mathop{y_1}\limits^\cdot = x_1 - a \\
\epsilon \mathop{x_2}\limits^\cdot = x_2 - {x_2}^3/3 - y_2 + I(\phi_1) + \rho(z) \cdot(x_1 - x_2) \\
\mathop{y_2}\limits^\cdot = x_2 - a \\
\mathop{z}\limits^\cdot = x_1 - x_2
\end{cases}.
\end{equation}
It is important to note that when $k_2=0$ term $\rho(z)(x_2-x_1)$ could be rewritten in the form $k_1 \cdot(x_2 - x_1)$ that corresponds to the most common way of description of electrical couplings (for details see, e.g. \cite{LevanovaPND}).

\subsection{Symmetries and first integrals}
\label{subsec:symmetries}

In this paper we consider identical elements and suppose that the couplings are symmetrical. Thus, system \eqref{ensembles} is invariant under change
\begin{equation}
\label{S_term}
S: x_1 \leftrightarrow x_2, y_1 \leftrightarrow y_2, z \leftrightarrow -z.
\end{equation}
Due to this symmetry for each trajectory passing through point $(x_1^*, y_1^*, x_2^*, y_2^*, z^*)$ there exists the trajectory passing through point $(x_2^*, y_2^*, x_1^*, y_1^*, -z^*)$ or this trajectory is self-symmetric. In particular, self-symmetric trajectories can belong to the invariant manifold
\begin{equation*}
\Pi: x_1 = x_2, y_1 = y_2, z = 0.
\end{equation*}
Such self-symmetrical solutions can be periodic and attractive. In this case they correspond to the in-phase neuron-like activity.

On the other hand, by definition, an anti-phase limit cycle containing a point $(x_1^*, y_1^*, x_2^*, y_2^*, z^*)$ should also pass through the point $(x_2^*, y_2^*, x_1^*, y_1^*, -z^*)$ after half a period. Thus, each anti-phase periodic cycle corresponds to a self-symmetrical solution that does not belong to invariant manifold $\Pi$. Moreover, the converse statement is also true: each self-symmetrical periodic regime of system \eqref{ensembles}, if it does not belong to invariant manifold $\Pi$, corresponds to an anti-phase limit cycle.

It is worth noting that, in addition to the symmetry \eqref{S_term}, system \eqref{ensembles} admits also the first integral \footnote{Indeed, it easy to see that $\dot{y_1} - \dot{y_2} = \dot{z}$. Integrating this expression one can obtain the first integral \eqref{1st_integral}}
\begin{equation}
\label{1st_integral}
F = y_1 - y_2 - z = const.
\end{equation}
On the level $F = C$ of this integral system \eqref{ensembles} defines the four-dimensional flow
\begin{equation} \label{ensembles_reduced}
\begin{cases}
\epsilon \mathop{x_1}\limits^\cdot = x_1 - {x_1}^3/3 - y_1 + I(\phi_2) + (k_1 + k_2 \cdot(y_1 - y_2 + C)^2)\cdot(x_2 - x_1)\\
\mathop{y_1}\limits^\cdot = x_1 - a\\
\epsilon \mathop{x_2}\limits^\cdot = x_2 - {x_2}^3/3 - y_2 + I(\phi_1) + (k_1 + k_2 \cdot(y_1 - y_2 + C)^2)\cdot(x_1 - x_2)\\
\mathop{y_2}\limits^\cdot = x_2 - a
\end{cases},
\end{equation}
where the value of the first integral $C$ can be considered as an additional governing parameter of the system, which gives additional opportunities to control neuron-like activity in the system.

In the next section \ref{sec:electrical_memristor_coupling} we show how parameters $k_1$, $k_2$ and $C$ affect the regimes of anti-phase, in-phase and sequential activity presented in figure~\ref{time_serieses}.

\section{Effects of electrical and memristor-based coupling}
\label{sec:electrical_memristor_coupling}
In this section we study the influence of electrical ($k_2 = 0$ in \eqref{ensembles_reduced}) and memristor-based ($k_2 \neq 0$) couplings to the regimes of in-phase $L_{in}$, anti-phase $L_{anti}$ and sequential ($L_{12}$ and $L_{1221}$) activity, see figure~\ref{time_serieses}, previously observed in ensemble \eqref{ensemble_chemical} with only chemical coupling.

First, we study the influence of electrical and memristor-based coupling to the regimes of in-phase activity. Recall, that all such regimes belong to the invariant manifold $\Pi: x_1 = x_2, y_1 = y_2$. The dynamics of system \eqref{ensembles_reduced} at this manifold is defined by the following system of two differential equations:
\begin{equation} \label{ensembles_PI}
\begin{cases}
\epsilon \mathop{x_1}\limits^\cdot = x_1 - {x_1}^3/3 - y_1 + I(\phi_2)\\
\mathop{y_1}\limits^\cdot = x_1 - a\\
\end{cases}.
\end{equation}
The detailed analysis of this system is presented in \cite{KorotkovCNSNS2019}, where, in particular, bifurcations leading to the appearance of the stable canard solutions \cite{Krupa2001} that correspond to the regimes of in-phase activity were studied.

Since parameters $k_1$ and $k_2$ are not included in system \eqref{ensembles_PI}, one can conclude that electrical and memristor-based couplings do not affect the dynamics on the invariant manifold $\Pi$ and, thus, coordinates of in-phase regimes do not change. But changes in $k_1$ and $k_2$ could affect to the stability of such regimes in the transversal to $\Pi$ direction. However, our numerical experiments show that it does not happen, i.e. the observed in-phase limit cycles remain stable with varying in both parameters $k_1$ and $k_2$.

On the other hand, the changes in $k_1$ and $k_2$ essentially affect the regimes of anti-phase $L_{anti}$ and sequential ($L_{12}$, $L_{1221}$) activity. In the following subsections we study bifurcations of this regimes in details.

\subsection{Chemical and electrical couplings ($k_2 = 0$)}
\label{sec:electrical_coupling}

In this subsection, we study how the increase in electrical coupling affects such type of neuron-like activity as anti-phase spiking regime $L_{anti}$ (see figure~\ref{time_serieses_a}) and regimes of sequential activity $L_{12}$ and $L_{1221}$ (see figures~\ref{time_serieses_c} and \ref{time_serieses_d}).

For this purpose we build bifurcation trees representing the dependency of coordinate $y_1$ of the steady-state regime on parameter $k_1$. Hereinafter the values of coordinate $y_1$ are taken on the Poincar\'e section $y_2 = 0$.

\subsubsection{Bifurcations of anti-phase spiking regime $L_{anti}$}

\begin{figure}[tbh]
	\centering
	\subfloat[]
	{
		\includegraphics[width = 0.33\linewidth]{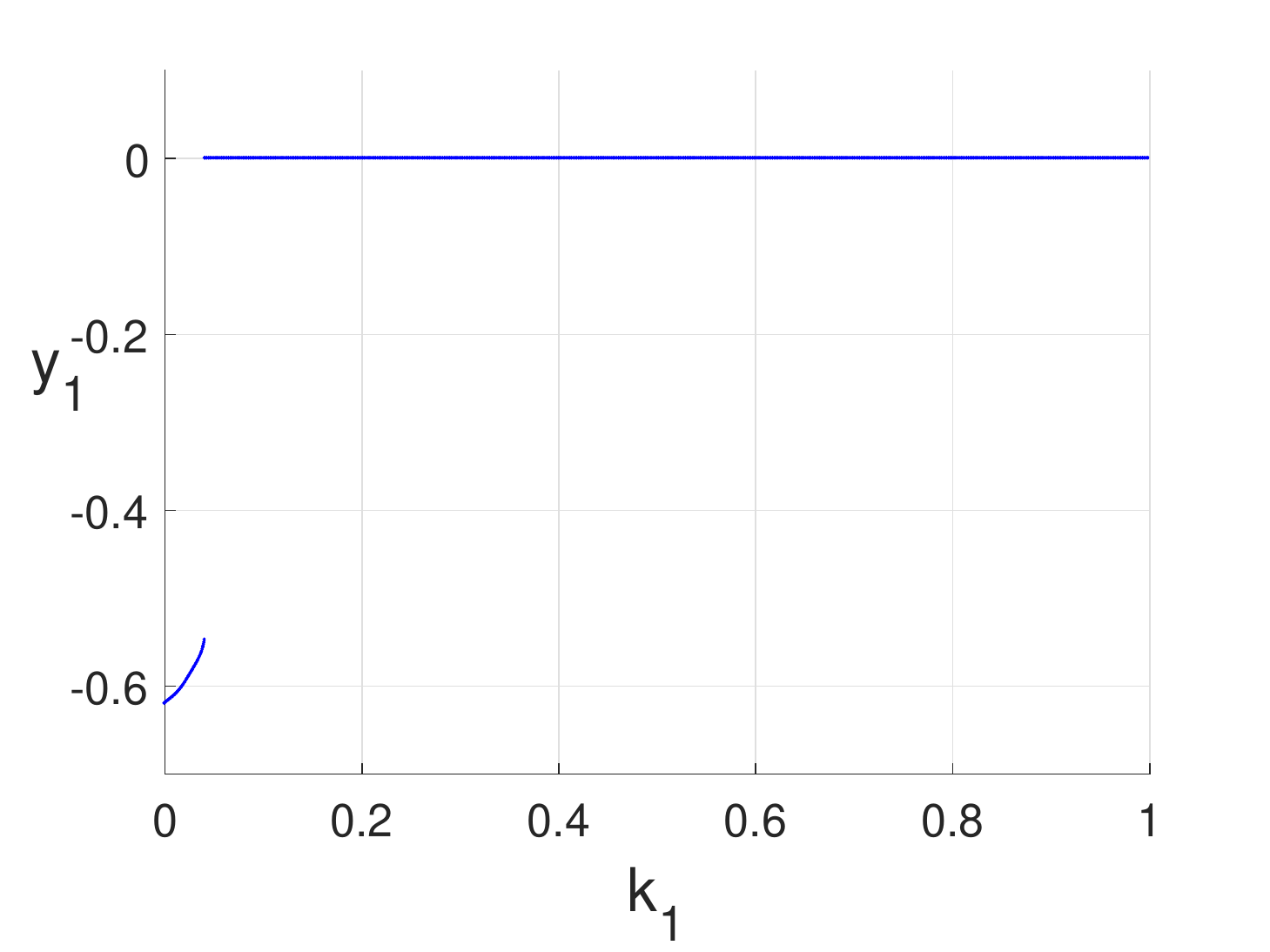}
		\label{Bif_tree_210_k2_0}
	}
	\subfloat[$k_1 = 0.02$]
	{
		\includegraphics[width = 0.33\linewidth]{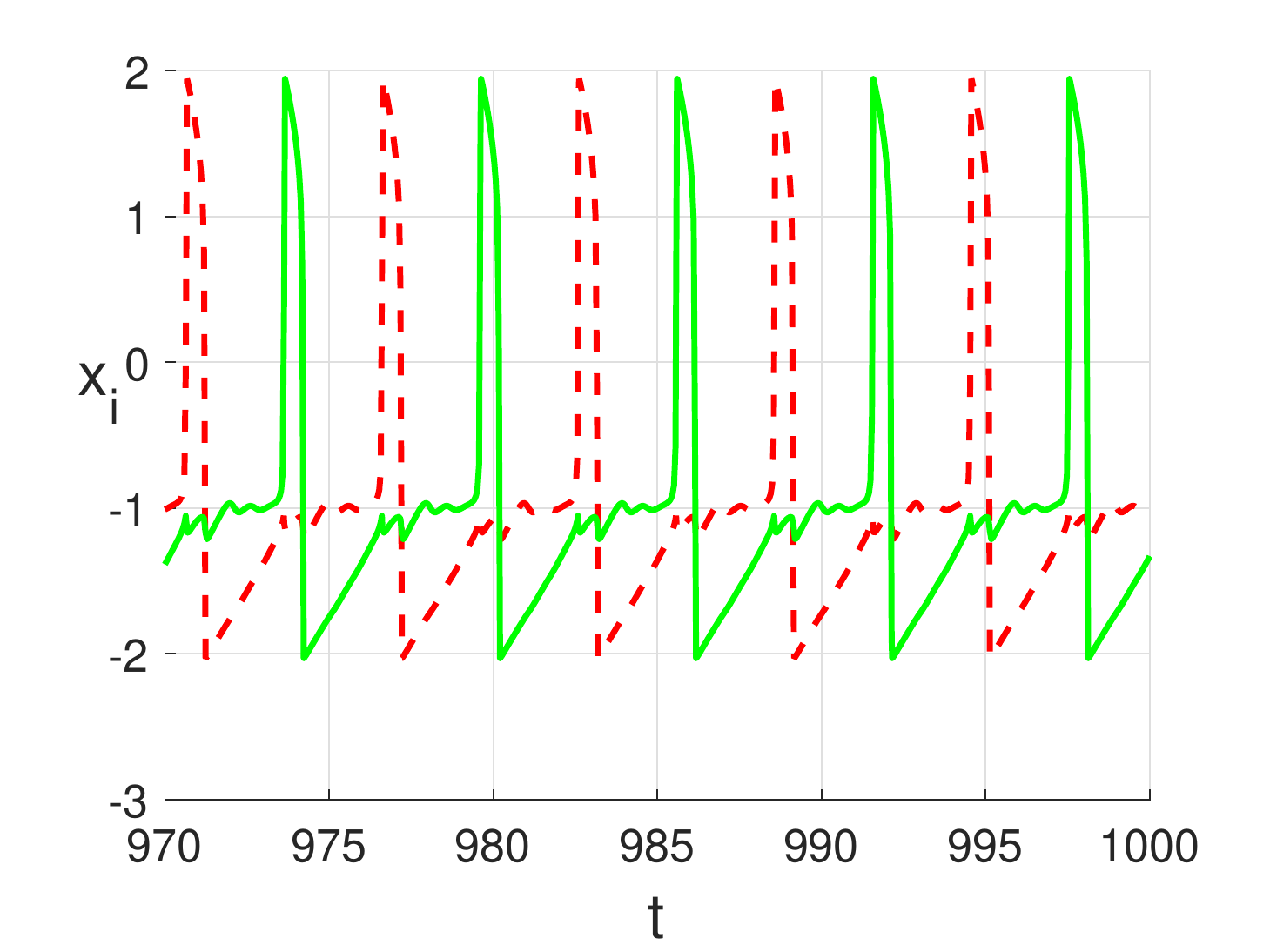}
		\label{TS_210_k1_0_02_k2_0}
	}
	\subfloat[$k_1 = 0.2$]
	{
		\includegraphics[width = 0.33\linewidth]{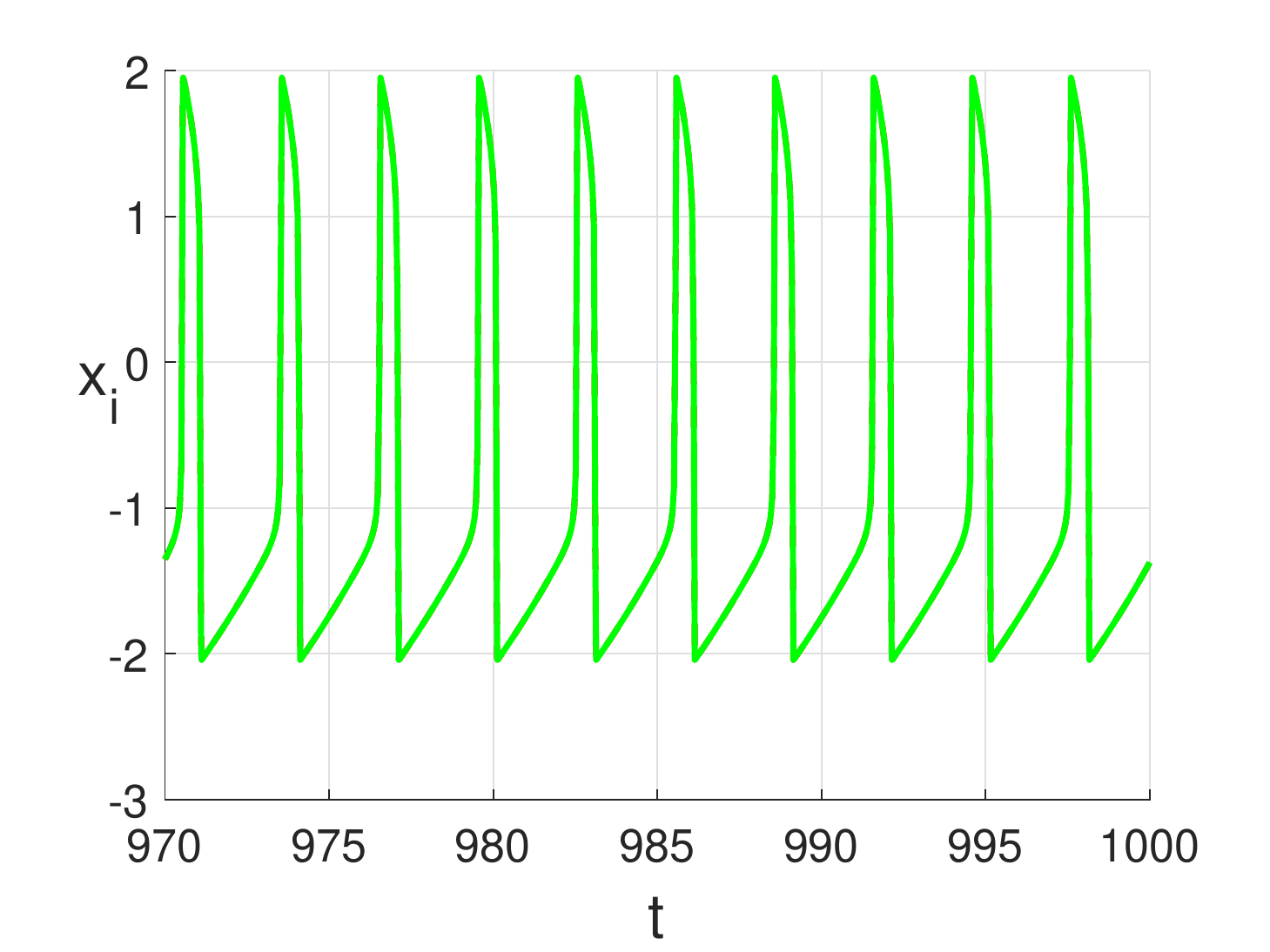}
		\label{TS_210_k1_0_2_k2_0}
	}
	\caption{Bifurcation tree and time series for cases $k_2 = 0$ of the system \eqref{ensembles_reduced} in the case of $\alpha = 210^\circ$.}
	\label{Bif_L_anti_k2_0}
\end{figure}

Figure~\ref{Bif_tree_210_k2_0} shows the bifurcation tree calculated for anti-phase spiking regime $L_{anti}$. As one can see, this limit cycle exists for $k_1 \leq k_{SN} \simeq 0.04$. At $k_1 = k_{SN}$ cycle $L_{anti}$ undergoes saddle-node bifurcations and, as result, orbits from its neighbourhood go to in-phase spiking regime $L_{in}$, see figure~\ref{TS_210_k1_0_2_k2_0}.

\subsubsection{Bifurcations of regime of sequential activity $L_{12}$}

\begin{figure}[tbh]
  \centering
  \subfloat[]
  {
    \includegraphics[width = 0.5\linewidth]{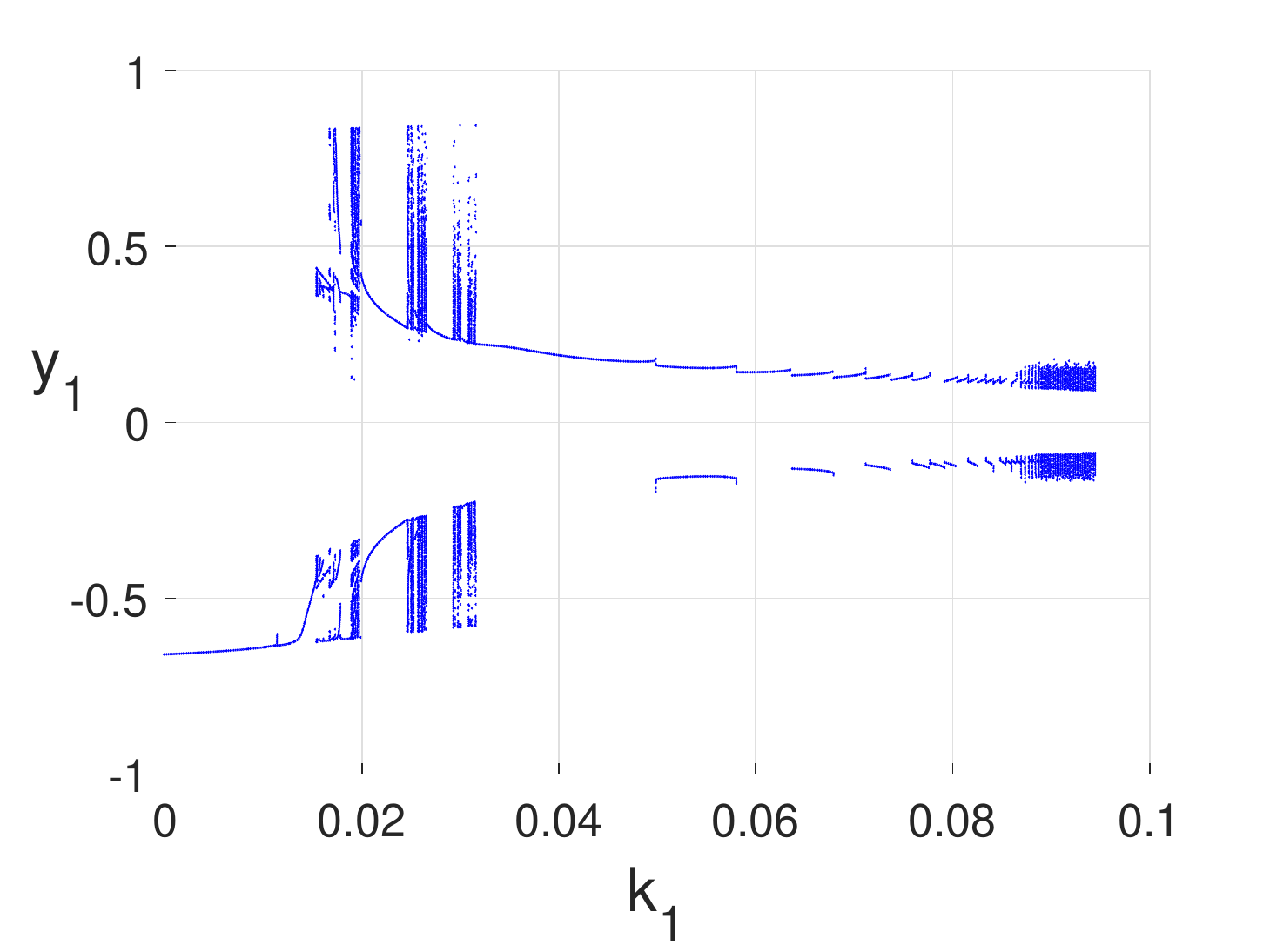}
    \label{bif_tree_160_k2_0}
  }\\
  \subfloat[$k_1 = 0.01$]
  {
    \includegraphics[width = 0.33\linewidth]{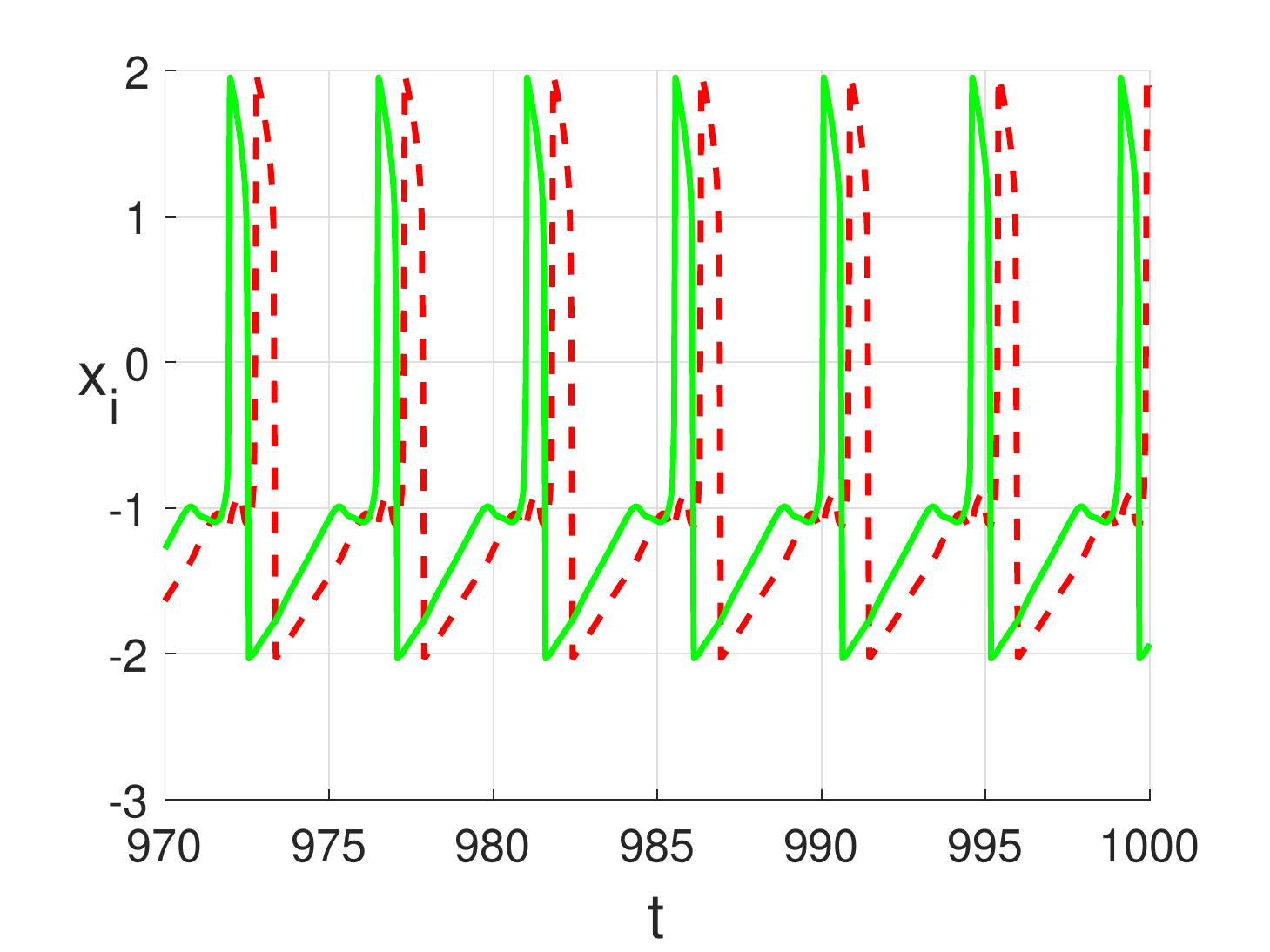}
    \label{TS_160_k1_0_01_k2_0}
  }
  \subfloat[$k_1 = 0.016$]
  {
    \includegraphics[width = 0.33\linewidth]{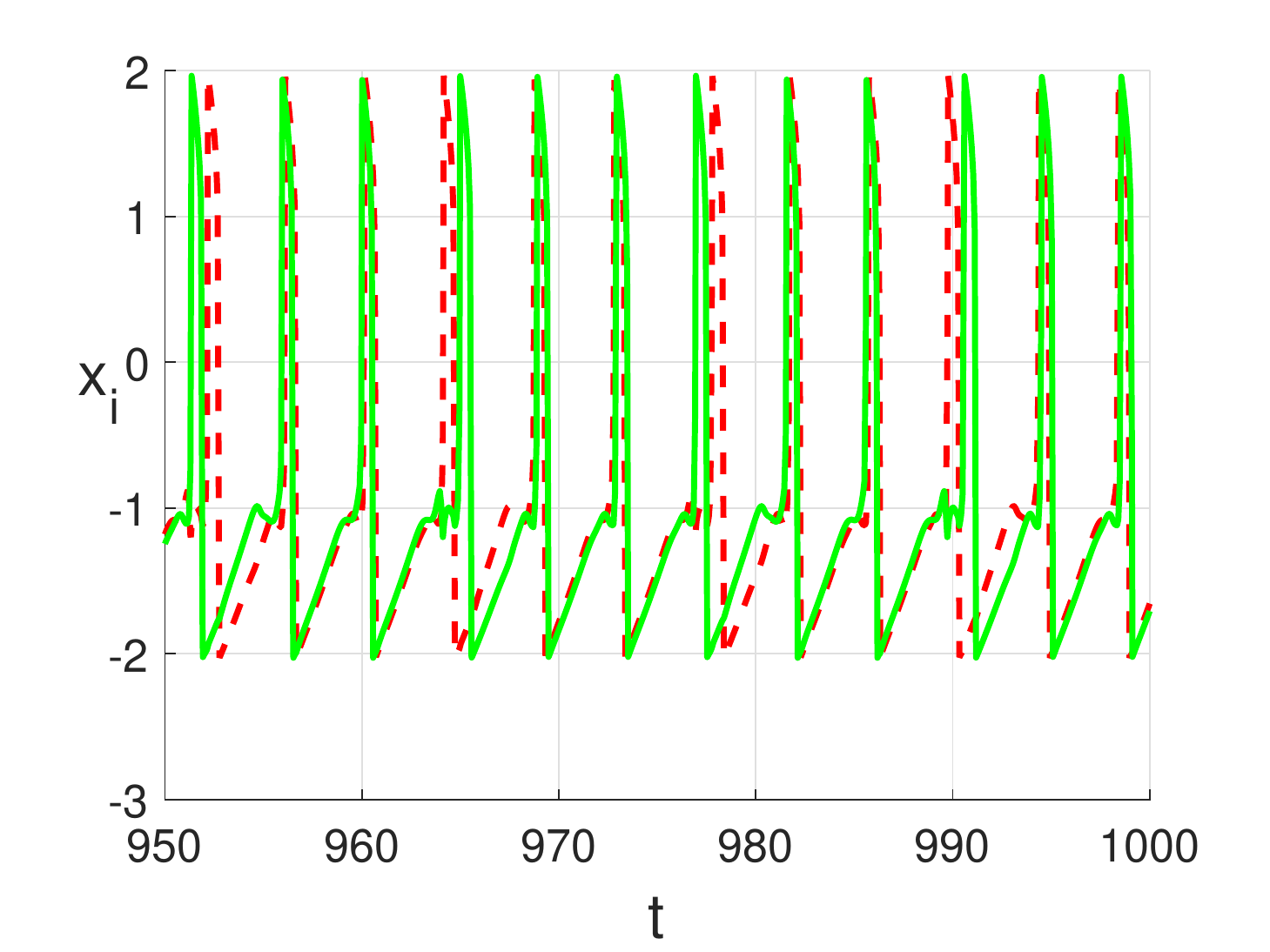}
    \label{TS_160_k1_0_016_k2_0}
  }
  \subfloat[$k_1 = 0.07$]
  {
    \includegraphics[width = 0.33\linewidth]{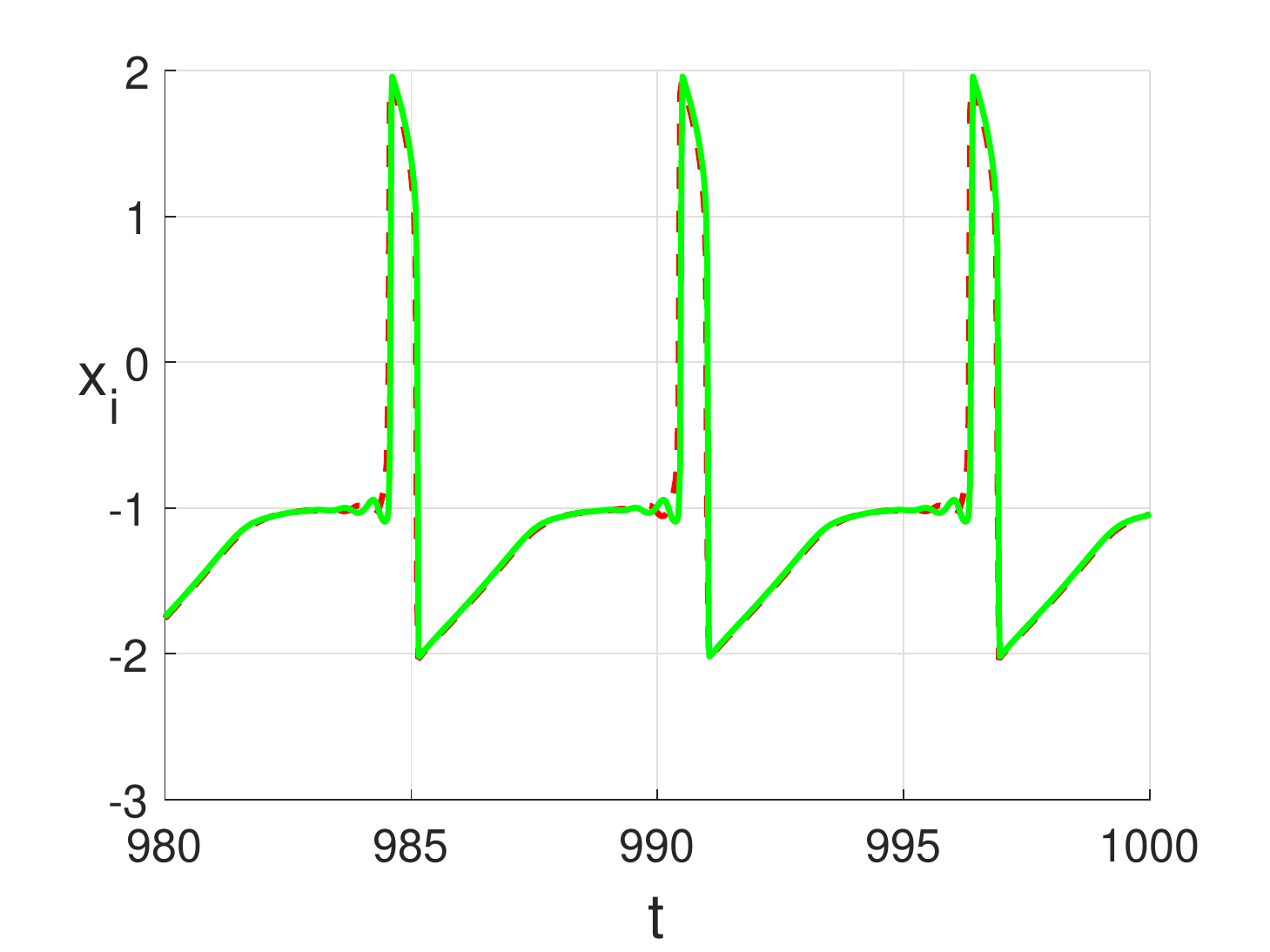}
    \label{TS_160_k1_0_07_k2_0}
  }
  \caption{Bifurcation tree and time series for cases $k_2 = 0$ of the system \eqref{ensembles_reduced} in the case of $\alpha = 160^\circ$.}
  \label{bif_160}
\end{figure}

With increasing in the electrical coupling parameter $k_1$ regimes of sequential activity $L_{12}$ and $L_{1221}$ undergo more complicated and interesting bifurcation scenarios. Here we show that even extreme events associated with arbitrary increasing of interspike intervals can occur here. Figure~\ref{bif_tree_160_k2_0} shows bifurcation tree for $L_{12}$. At small values $k_1$ regime $L_{12}$ persists, see time series in figure~\ref{TS_160_k1_0_01_k2_0}. With increasing in $k_1$, bifurcations of this regime give rise to chaotic behaviour. From figure~\ref{bif_tree_160_k2_0} one can see that regions with chaotic behaviour alternate with the so-called stability windows, inside which stable periodic orbits with different period appear. Note that such periodic orbits may correspond to different types of sequential neuron-like activity (see e.g. figure~\ref{TS_160_k1_0_016_k2_0} and \ref{TS_160_k1_0_07_k2_0}), where regimes with the following sequences of activated elements occur.

Let us now study the evolution of chaotic behaviour. Few bifurcations scenarios leading to the appearance of chaotic attractors associated with regime $L_{12}$ are observed in the system \eqref{ensembles_reduced}. The first scenario is the cascade of period-doubling bifurcations. This type of transition to chaos is observed e.g. at $k_1 \approx 0.03115$.

\begin{figure}[tbh]
	\centering
		\subfloat[]
	{
		\includegraphics[width = 0.5\linewidth]{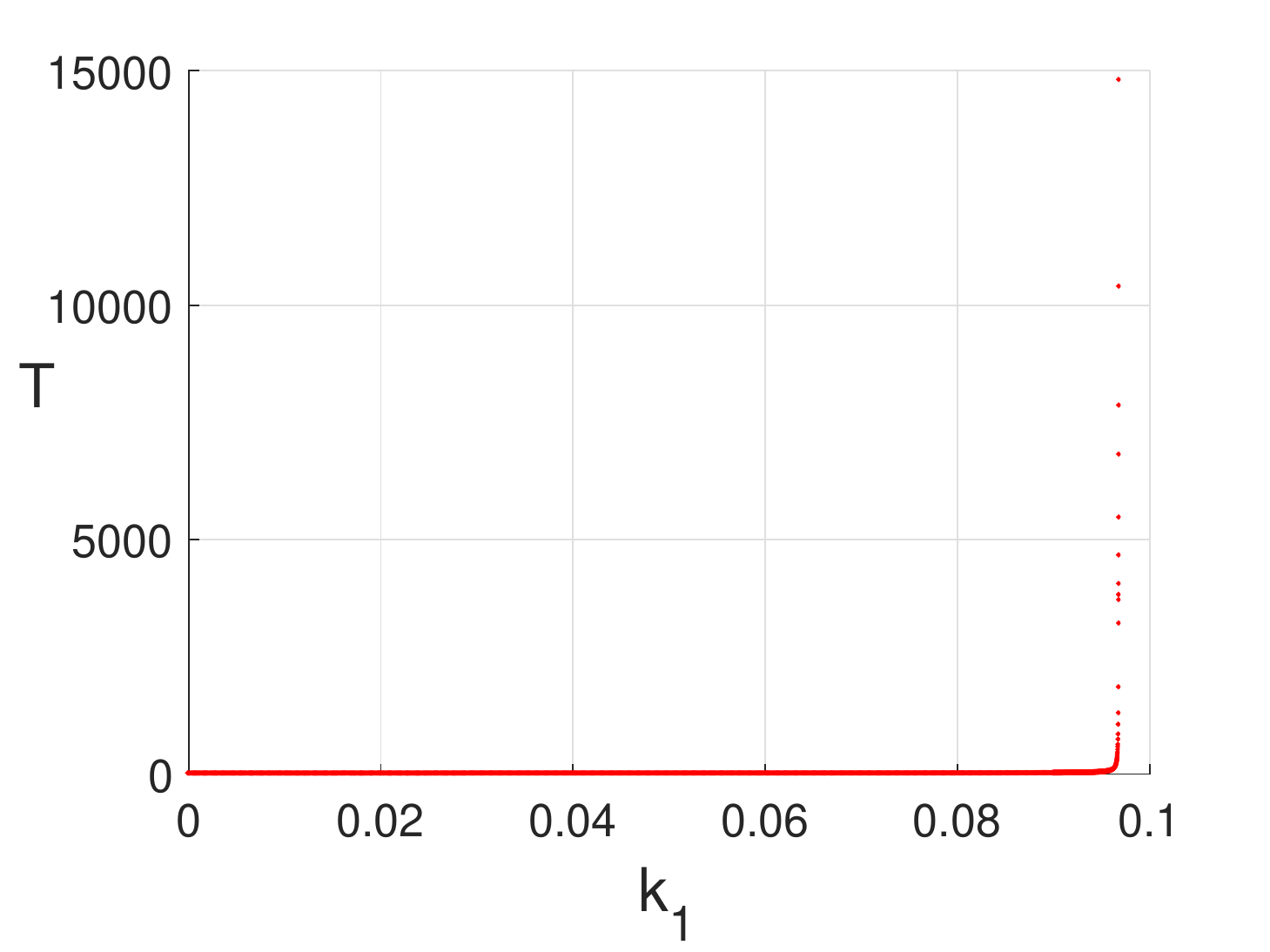}
		\label{period_homoclinic}
	}
	\subfloat[]
	{
		\includegraphics[width = 0.5\linewidth]{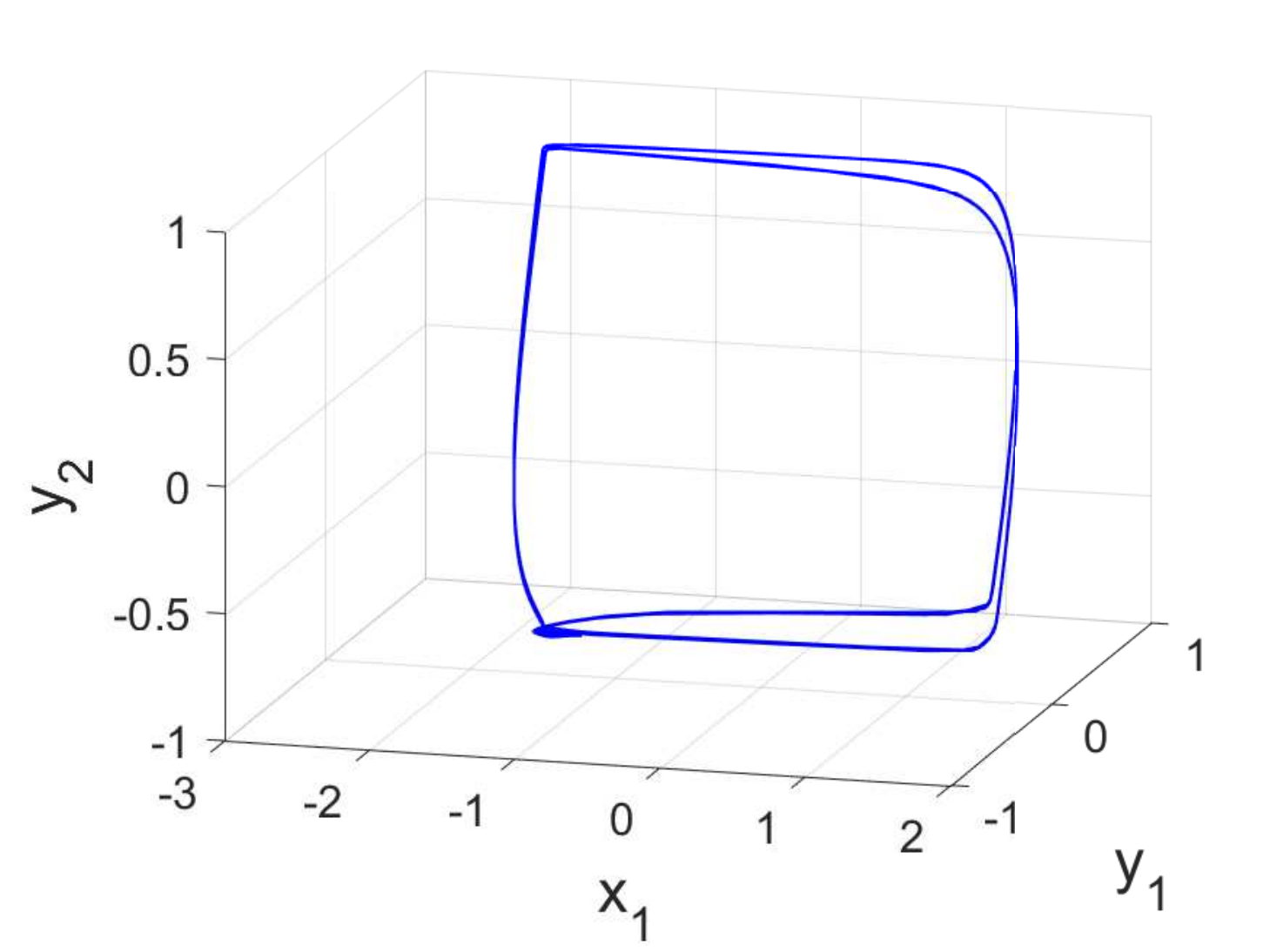}
		\label{PP_homoclinic_b}
	}\\
	\subfloat[]
	{
		\includegraphics[width = 0.5\linewidth]{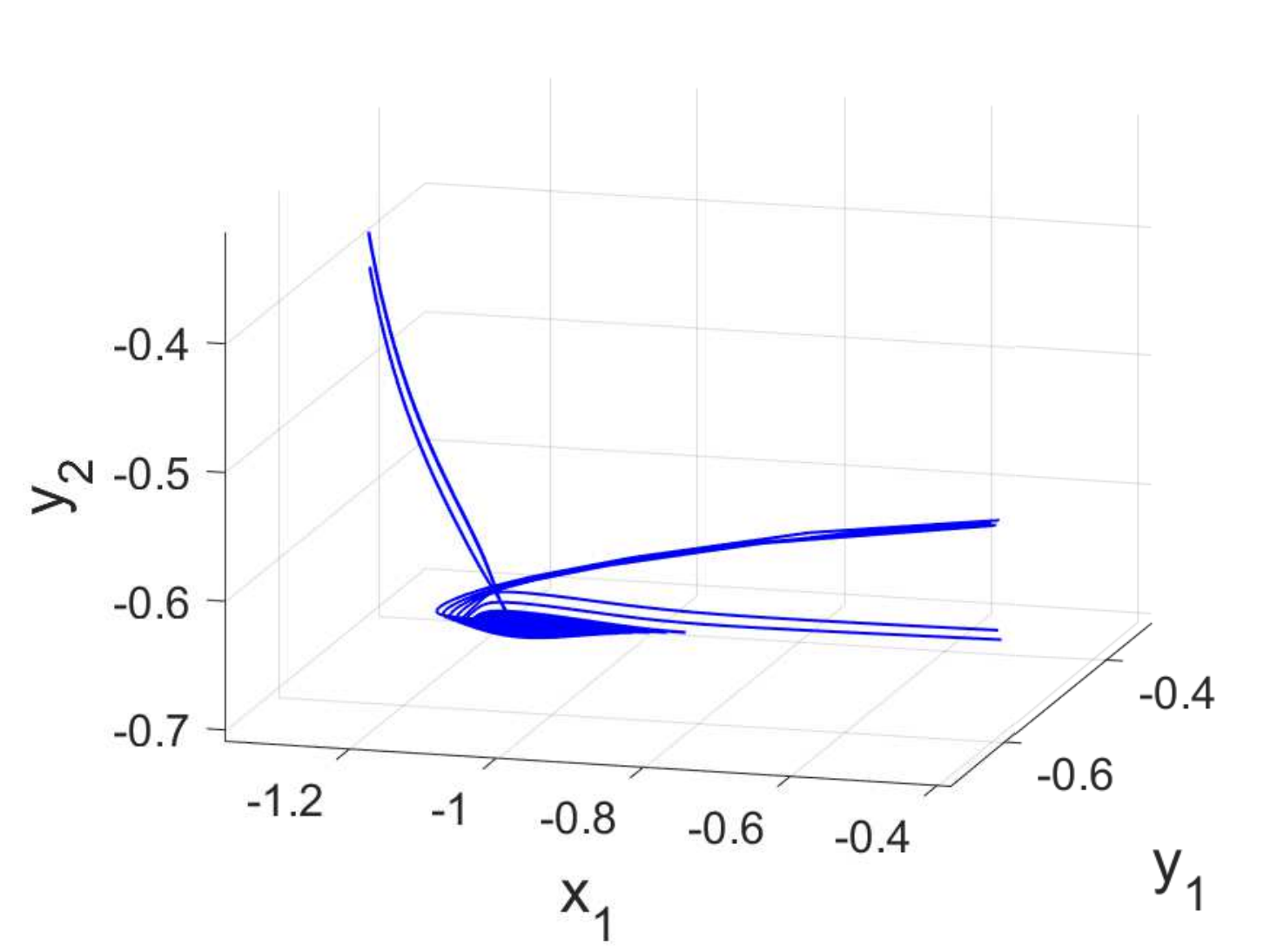}
		\label{PP_homoclinic_c}
	}
	\subfloat[]
	{
		\includegraphics[width = 0.5\linewidth]{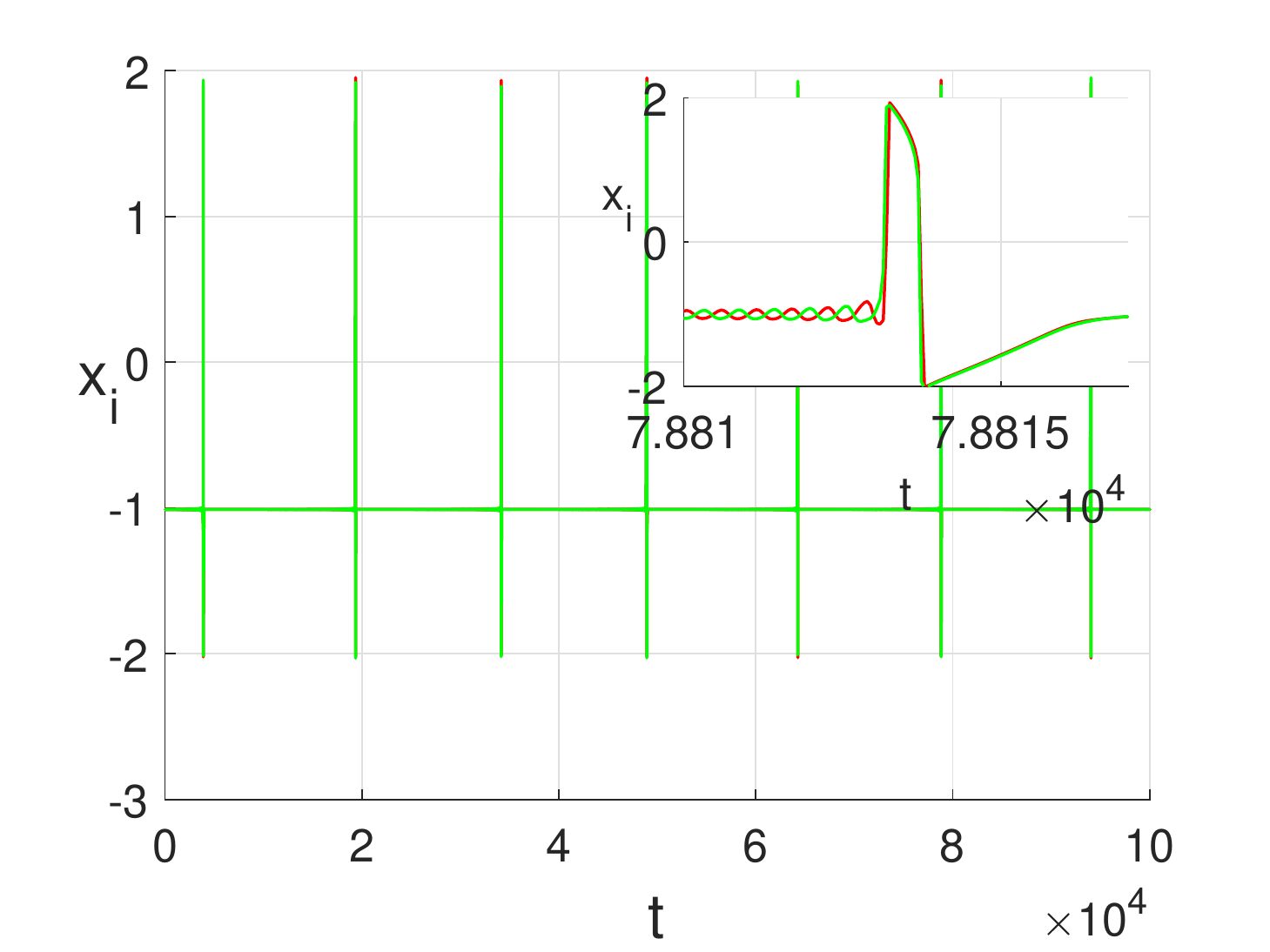}
		\label{TS_homoclinic}
	}
	\caption{(a) The dependence of interspike interval on $k_1$. (b) Strange spiral attractor containing saddle-focus equilibrium with its homoclinic orbit, and (c) its enlarged part and (d) time series. Parameters' values: $\alpha = 160^\circ$, $k_1=0.096858$.}
	\label{PP_homoclinic}
\end{figure}

The second and the most interesting scenario is associated with Shilnikov bifurcation \cite{Shil65,Shil70} and observed for $k_1 = {k_1}^* \simeq 0.097$. Due to this bifurcation strange spiral attractor containing saddle-focus equilibrium with its homoclinic orbit emerges in the phase space of system \eqref{ensembles_reduced} (see figure~\ref{PP_homoclinic}). Interspike intervals become elongated when $k_1$ tends to ${k_1}^*$, see figure~\ref{period_homoclinic}. Finally, at $k_1 = {k_1}^*$, homoclinic spiral attractor appears, see figure~\ref{PP_homoclinic_b}, \ref{PP_homoclinic_c}. Note that this chaotic regime corresponds to quasi-in-phase chaotic neuron-like activity (the states of both elements are approximately equal), see figure~\ref{TS_homoclinic}.

It is important to note, that the emergence of the described above regime can be considered as an extreme event \cite{extreme_event}. Since the saddle-focus equilibrium belongs to the observed spiral attractor, orbits of this attractor can pass arbitrary close to this equilibrium. Thus interspike intervals can be arbitrarily large, what may mean the total suppression of the neuron.

A small change in the electrical coupling parameter $k_1$ gives opportunity to control this phenomenon. Indeed, arbitrarily small increase in $k_1$ destroys homoclinic orbit of the saddle-focus. However, other (the so-called secondary and multi-round) homoclinic orbits can appear here \cite{GasGonTur}. Besides, interspike intervals can remain sufficiently large for small change in $k_1$ even when the attractor does not contain the saddle-focus.

On the other hand, further increase in $k_1$ (when $k_1 > {k_1}^*$) leads to the crisis of strange attractor associated with regime $L_{12}$, after which almost all orbits in its neighborhood pass to the stable equilibrium. Thus, we can conclude that the further increase in the value of parameter of electrical couplings not only destroys homoclinic orbits, but also abolishes chaotic regimes associated with $L_{12}$.

However, quasi-in-phase neuron-like regime with quite large interspike intervals can simulate some real processes in neural ensembles \cite{long_interspike}. Also, as was shown in \cite{BakKazKorLevOs18}, such regimes can be both regular and chaotic. In the next section we show that taking into account the coupling through the electromagnetic field based on a memristor give a possibility to control such regimes. In particular, such coupling can help to avoid extreme events associated with the emergence of homoclinic spiral attractors by transforming them to non-homoclinic or even regular attractive sets.

\subsection{Chemical and memristor-based couplings ($k_2 \neq 0$)}
\label{sec:memristive_coupling}
 
In this subsection we study the impact of memristor-based coupling to the regimes of anti-phase activity $L_{anti}$, sequential activity $L_{12}$ and $L_{1221}$, and chaotic quasi-in-phase activity corresponding to the homoclinic spiral attractor described in the previous subsection.
 
\begin{figure}[h!]
  \centering
  \subfloat[$\alpha = 210^\circ$]
  {
    \includegraphics[width = .33\linewidth]{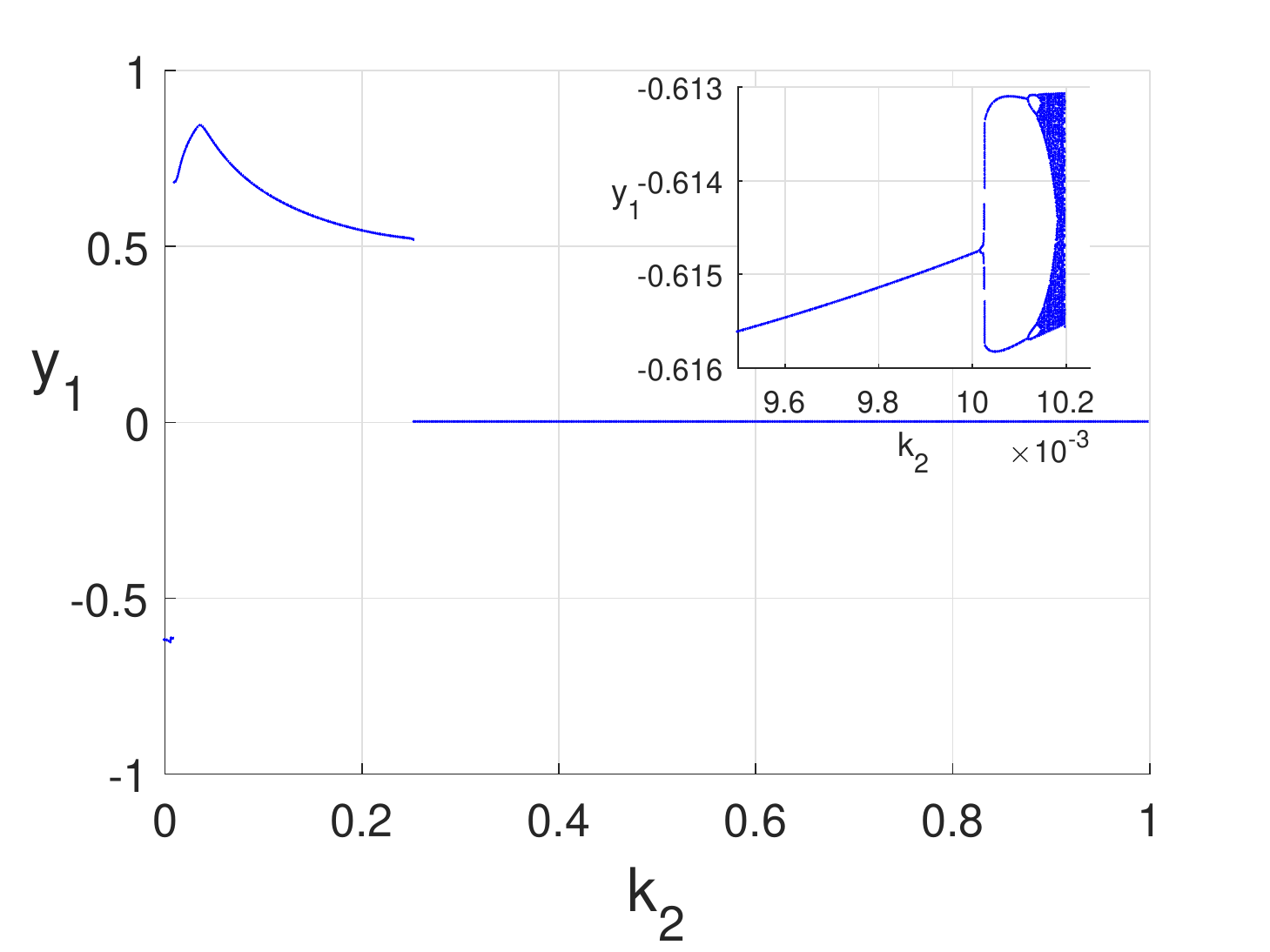}
    \label{bif_tree_210_k1_0}
  }
  \subfloat[$\alpha = 210^\circ$, $k_2 = 0.2$]
  {
    \includegraphics[width = .33\linewidth]{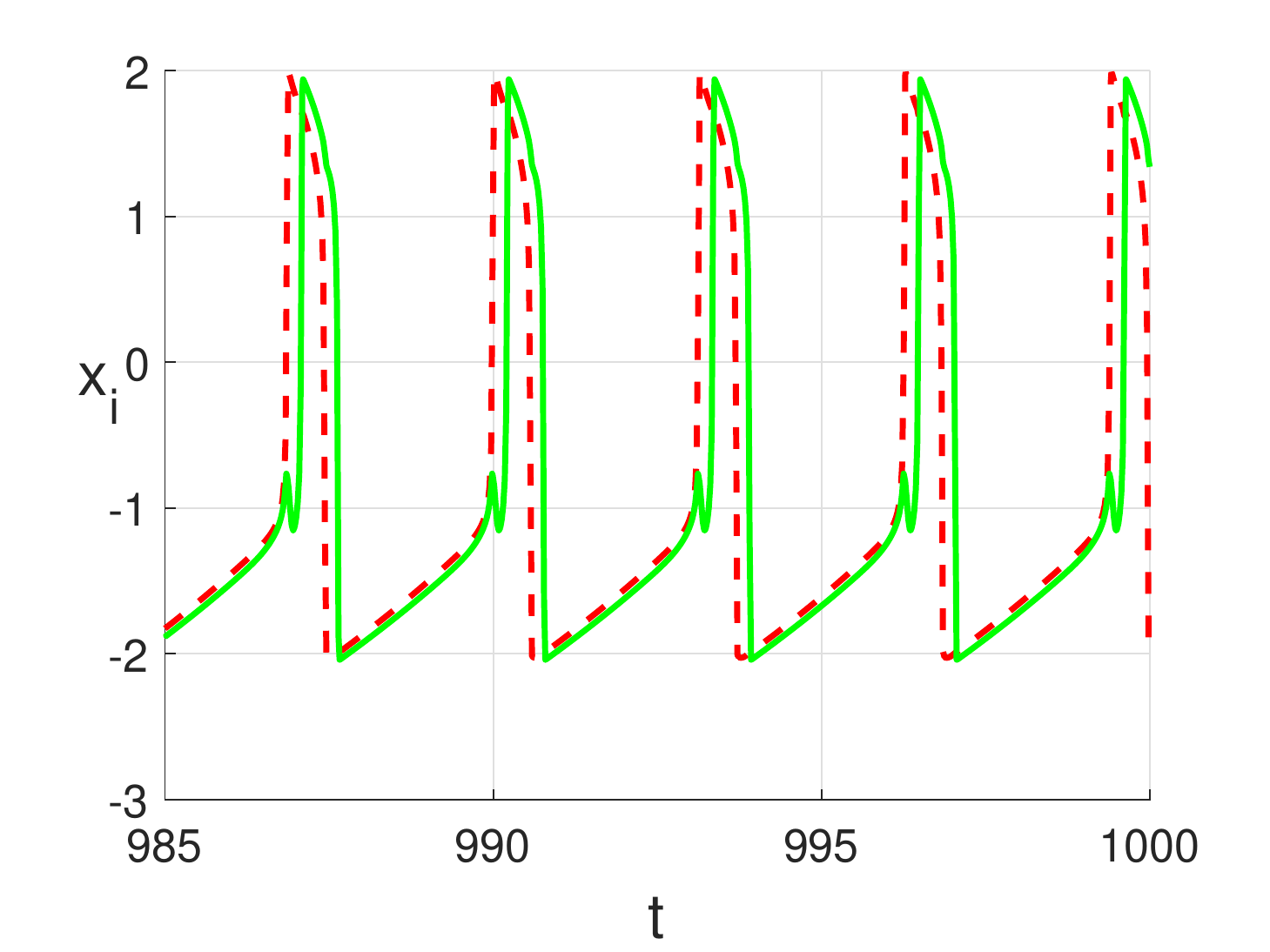}
    \label{TS_210_k1_0}
  }
  \subfloat[$\alpha = 160^\circ$]
  {
    \includegraphics[width = .33\linewidth]{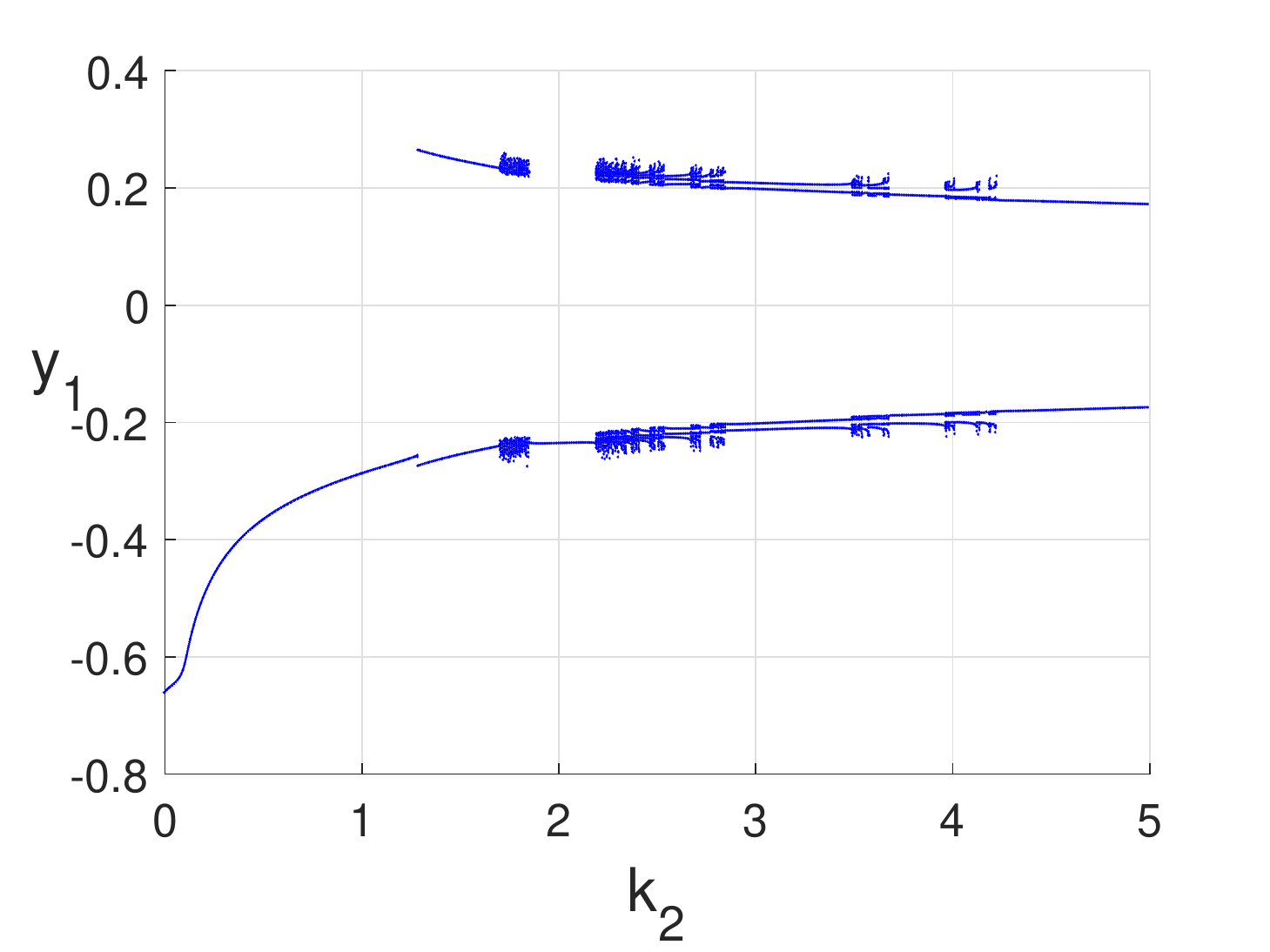}
    \label{bif_tree_160_k1_0}
  }\\
  \subfloat[$\alpha = 160^\circ$, $k_2 = 0.8$]
  {
    \includegraphics[width = .33\linewidth]{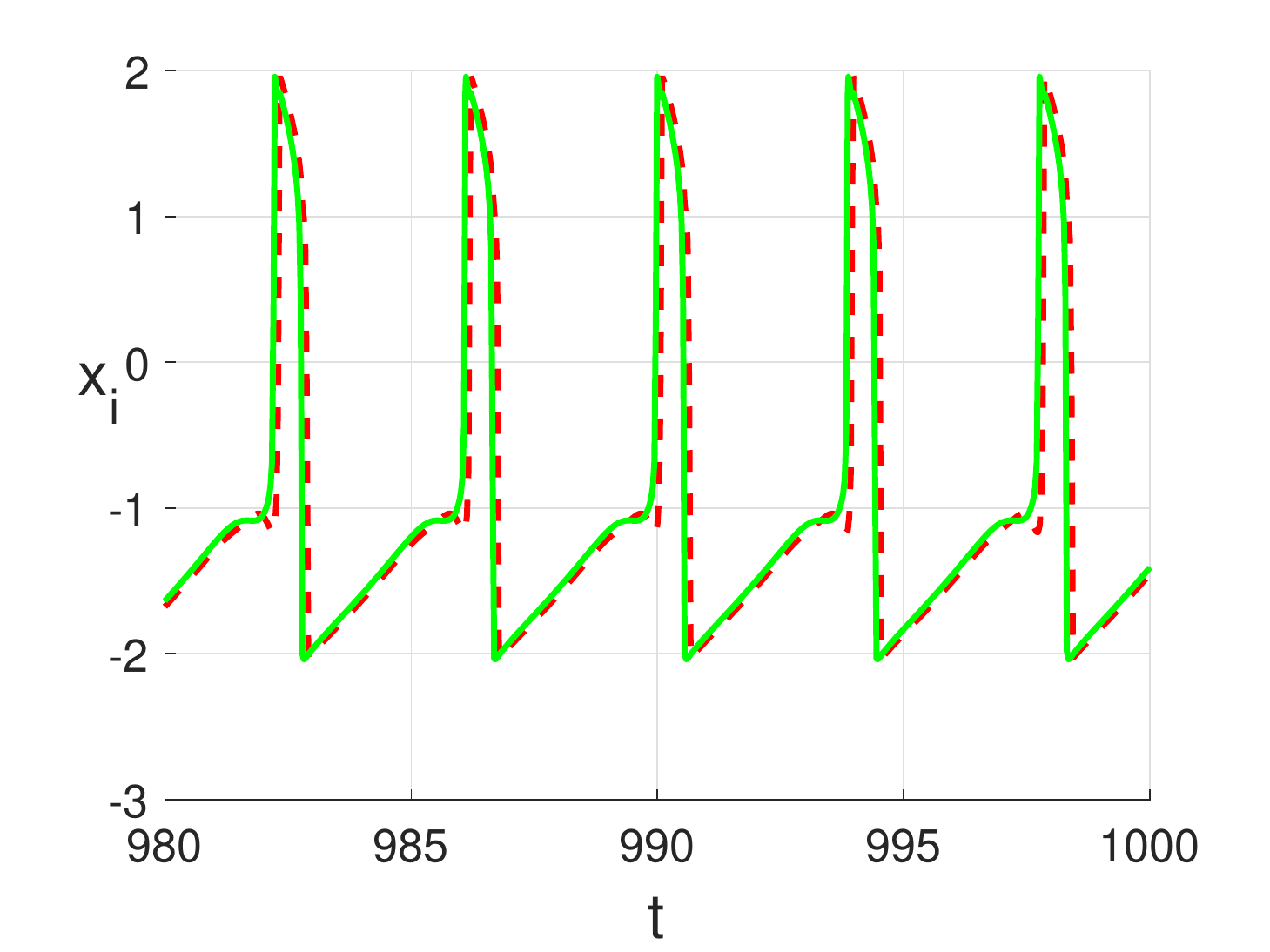}
    \label{TS_160_k1_0}
  }
  \subfloat[$\alpha = 159^\circ$]
  {
    \includegraphics[width = .33\linewidth]{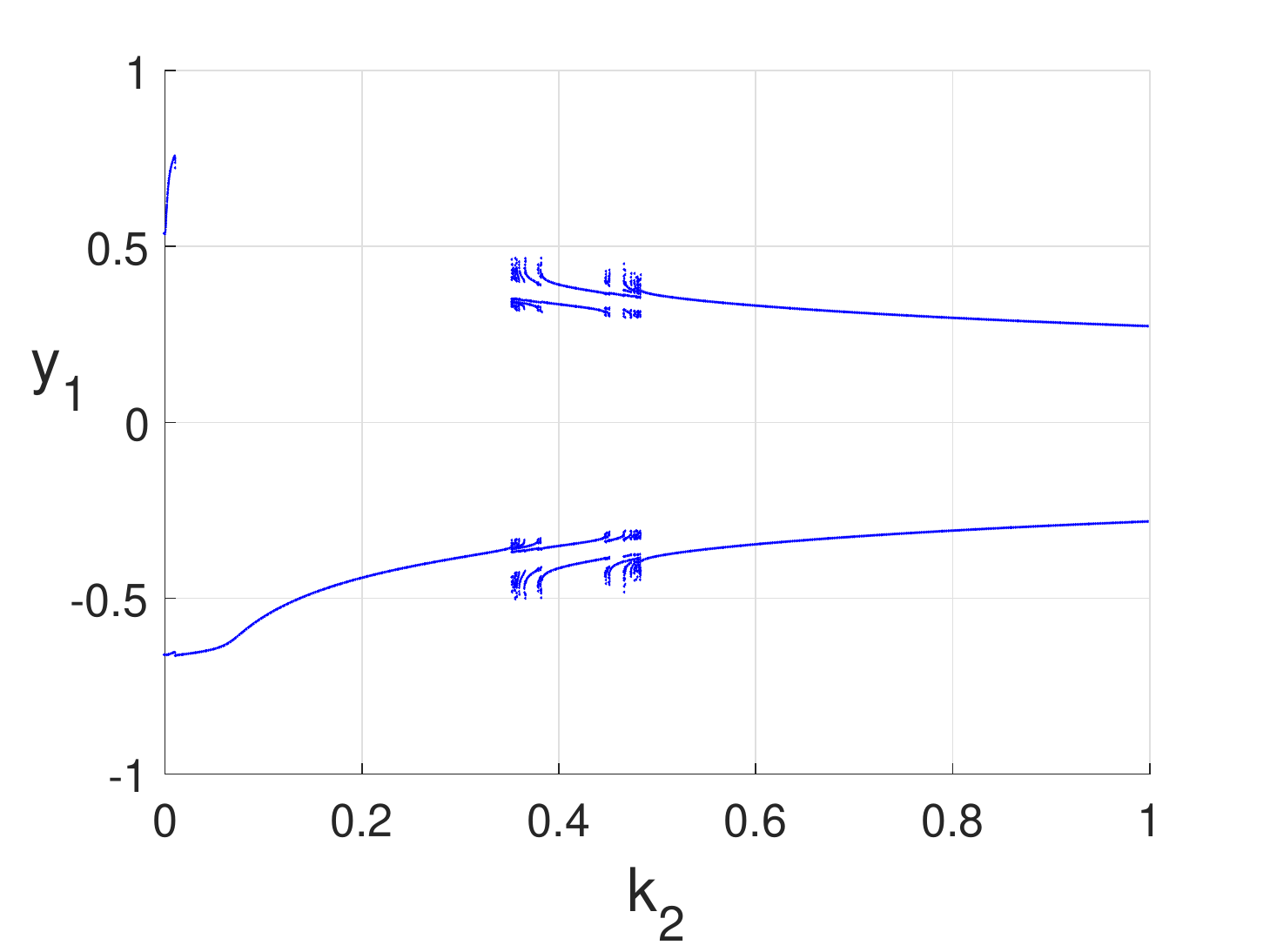}
    \label{bif_tree_159_k1_0}
  }
  \subfloat[$\alpha = 159^\circ$, $k_2 = 0.2$]
  {
    \includegraphics[width = .33\linewidth]{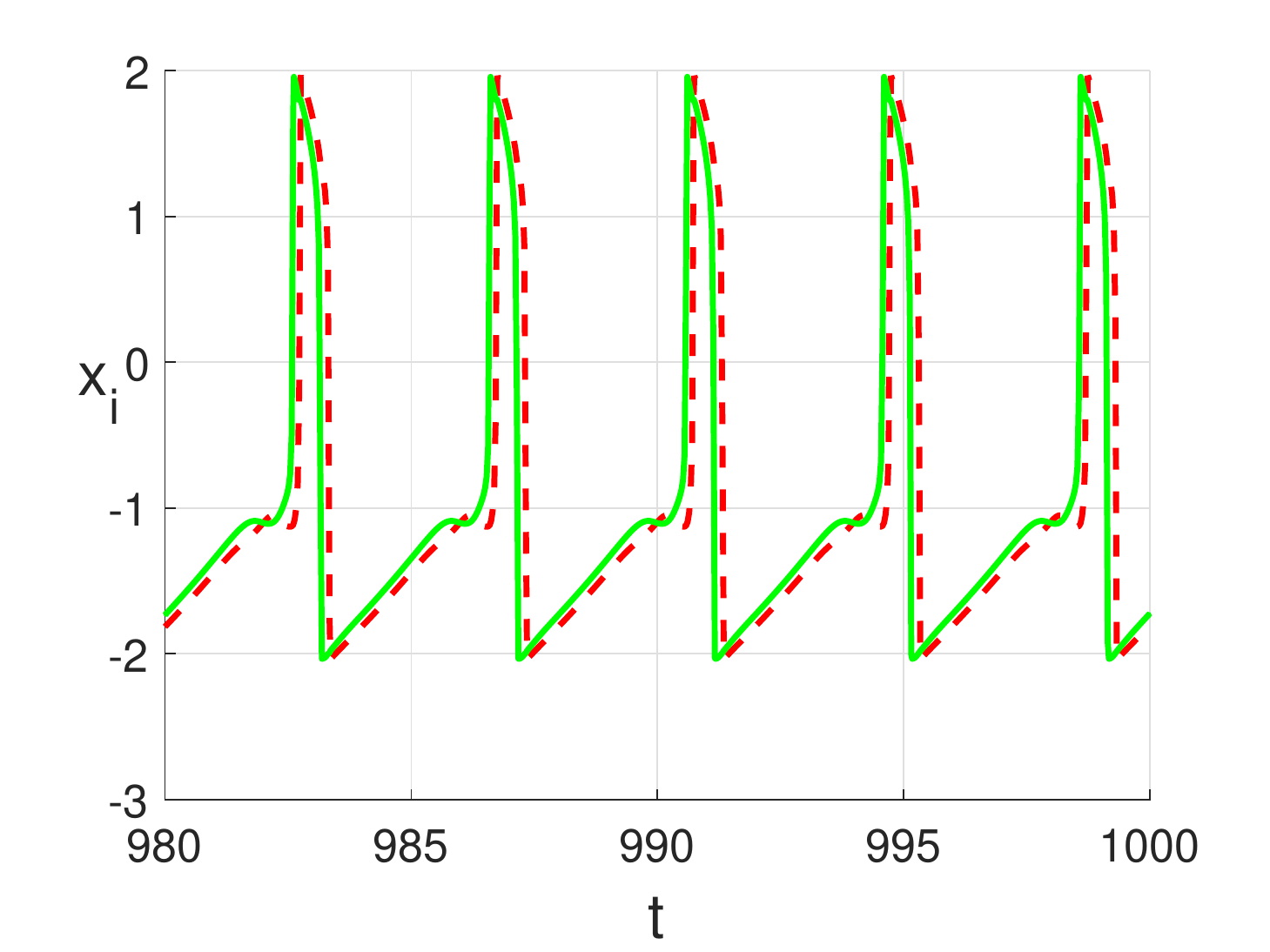}
    \label{TS_159_k1_0}
  }
  \caption{The impact of $k_2$ parameter on $L_{1221}$, $L_{12}$ and $L_{anti}$ cycles while $k_1 = 0$}
  \label{bif_trees_k2_var}
\end{figure}

The bifurcation trees illustrating the impact of $k_2$ to the regimes $L_{anti}$, $L_{12}$ and $L_{1221}$ are presented in the upper row in figure~\ref{bif_trees_k2_var}. Time series for the regimes developing from $L_{anti}$, $L_{12}$ and $L_{1221}$ are presented in the bottom row in figure~\ref{bif_trees_k2_var}.
 
Figure~\ref{bif_tree_210_k1_0} shows the bifurcation tree for $L_{anti}$. At $k_2 \in [0.01, 0.0102]$ this cycle undergoes the cascade of period doubling bifurcations and, as a result, Feigenbaum-like attractor appears (see the corresponding time series in figure~\ref{TS_210_k1_0}). At $k_2 \approx 0.0102$, the described attractor undergoes crisis, after which trajectories from its neighbourhood pass to another limit cycle, which corresponds to the regime of in-phase activity.
 
The bifurcation tree for the regime of sequential activity $L_{12}$ is presented in figure~\ref{bif_tree_160_k1_0}. In contrast to the anti-phase limit cycle, $L_{12}$ depends smoothly on $k_2$. The first element always activates the second one. However, time interval required to transmit signal between these elements is decreased with increasing in $k_2$, i.e. this regime smoothly tends to quasi-in-phase limit cycle, see figure~\ref{TS_160_k1_0}.
 
The impact of increasing the coupling parameter $k_2$ on the regime of sequential activity $L_{1221}$ is similar to the impact of increasing $k_2$ on $L_{12}$, see figure~\ref{bif_tree_159_k1_0}. However, in this case the coordinate $y_1$ of $L_{1221}$ does not depend continuously on the coupling parameter $k_2$. At $k_2 \in [0.35, 0.5]$ regime developed from $L_{1221}$ undergoes several bifurcations, after which a regime of sequential activity of the same type appears again, see figure~\ref{TS_159_k1_0}.  As for regime $L_{12}$, time required to transmit a signal from synaptic element to presynapic is decreased with increasing in $k_2$. Thus, both regimes of sequential activity tends to quasi-in-phase regimes with increasing of the strength of memristor-based coupling.
 
\subsubsection{Bifurcations of the regime of chaotic quasi-in-phase activity}
 
In section~\ref{sec:electrical_coupling} it was shown that increasing the electrical coupling parameter $k_1$ leads to the appearance of homoclinic spiral attractors $L_{hom}$ from the regimes of sequential activity $L_{12}$ and $L_{1221}$. Such attractors correspond to an interesting regimes of chaotic quasi-in-phase spiking activity for which interspike interval can be arbitrarily large. Further we demonstrate how memristor-based coupling affects to these regimes.
 
First we note that since the observed homoclinic attractors correspond to the regimes similar to in-phase activity, the value of term $y_1 - y_2 + C$ in system \eqref{ensembles_reduced} close to the value of parameter $C$. Thus, for $C = 0$ parameter $k_2$ weakly affects to these regimes. 
 
\begin{figure}[h!]
  \centering
  \subfloat[]
  {
    \includegraphics[width = 0.6\linewidth]{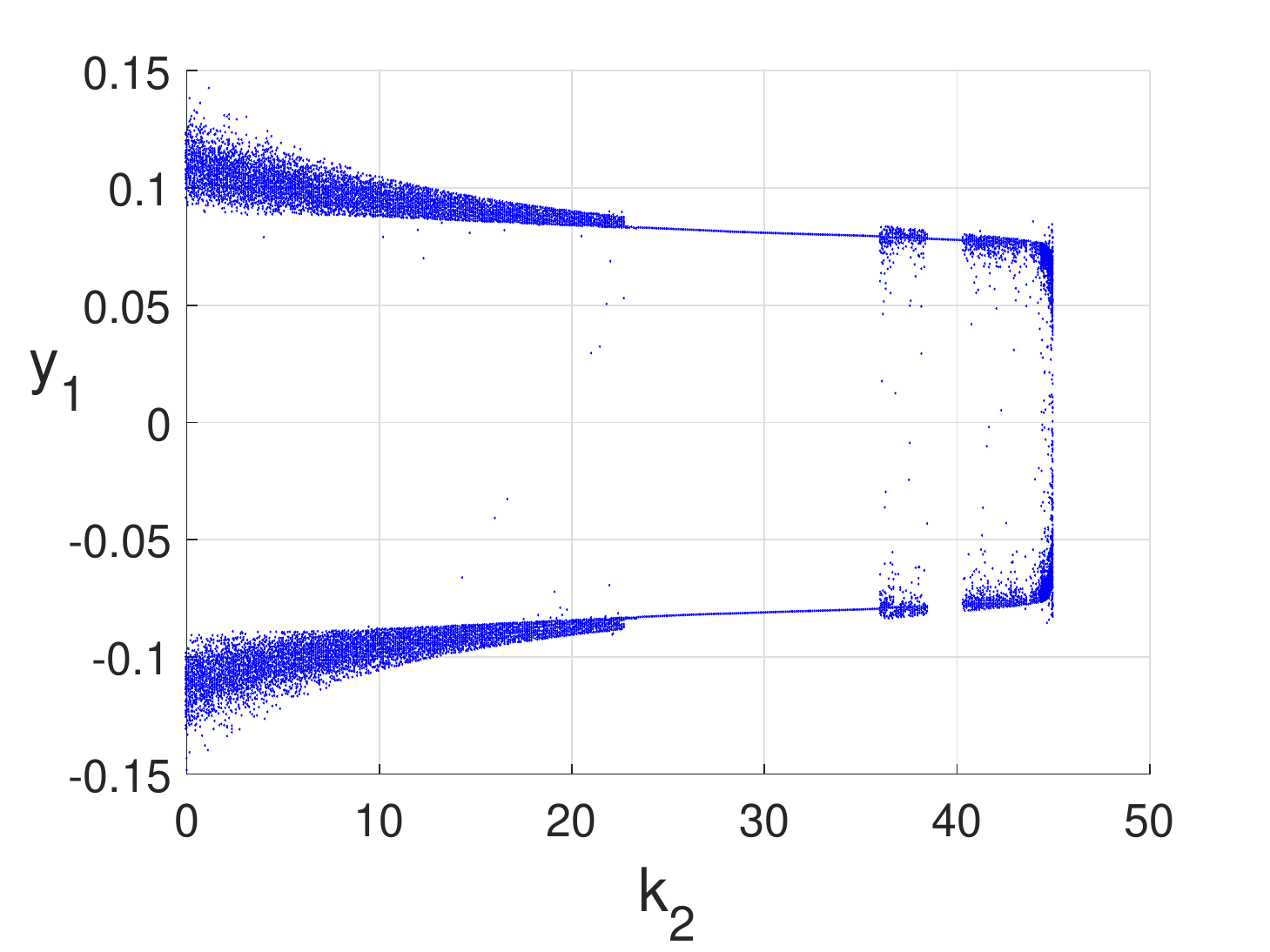}
    \label{bif_tree_160_k1_0_094589}
  }\\
  \subfloat[$k_2 = 45$]
  {
    \includegraphics[width = 0.33\linewidth]{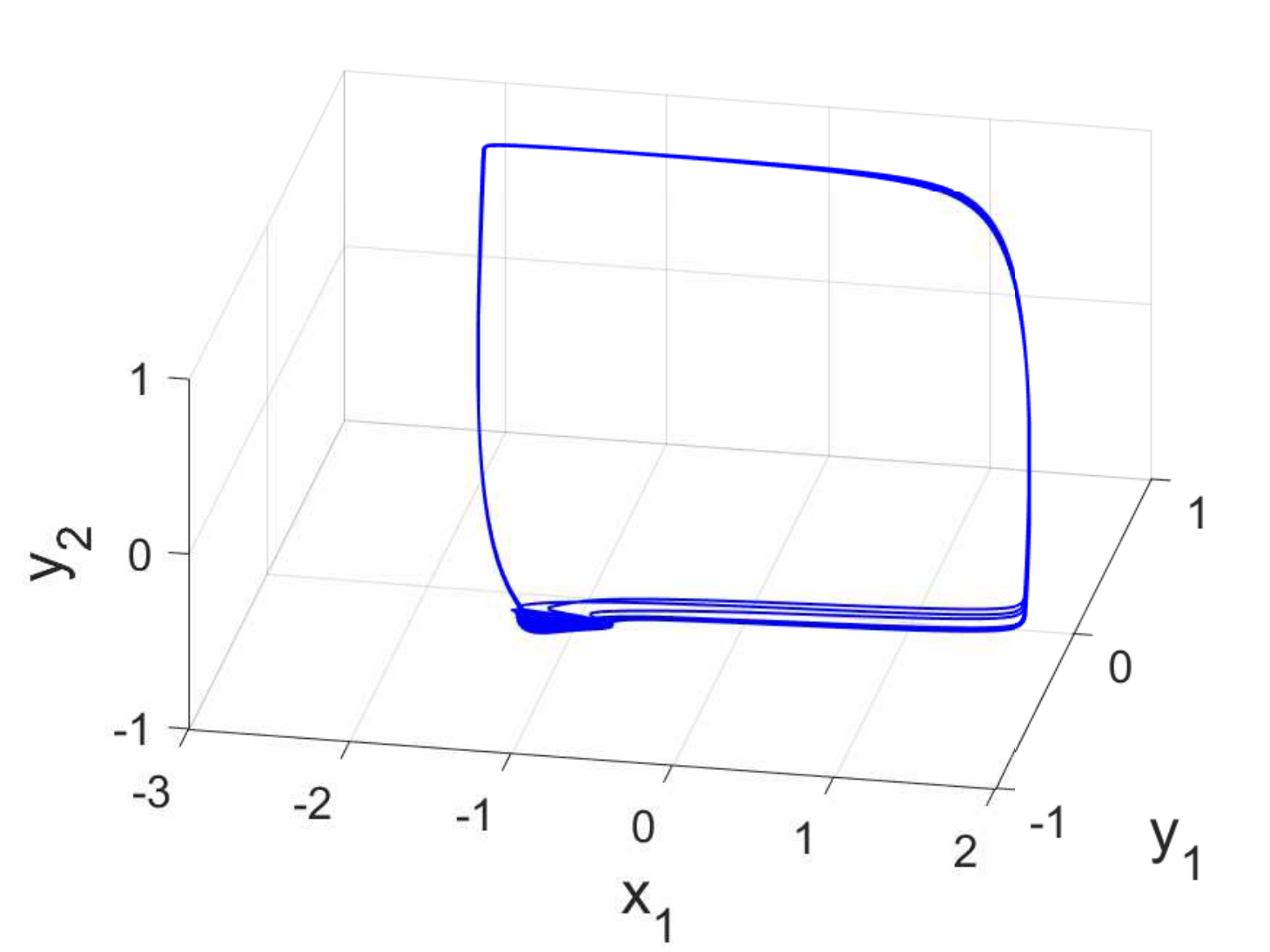}
    \label{PP_160_k1_0_094589_k2_45}
  }
  \subfloat[$k_2 = 45$]
  {
    \includegraphics[width = 0.33\linewidth]{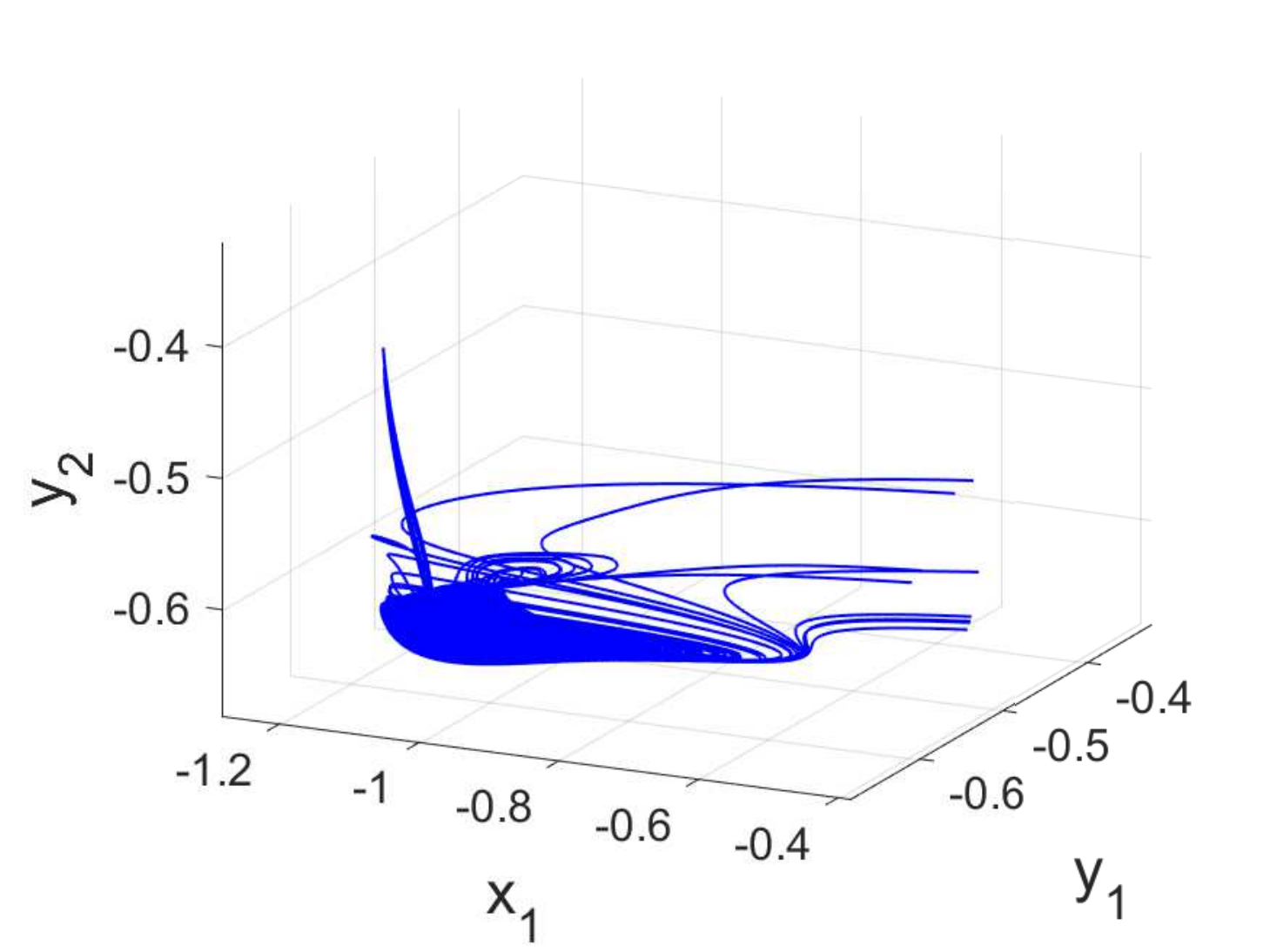}
    \label{PP_160_k1_0_094589_k2_45_part}
  }
  \subfloat[$k_2 = 45$]
  {
    \includegraphics[width = 0.33\linewidth]{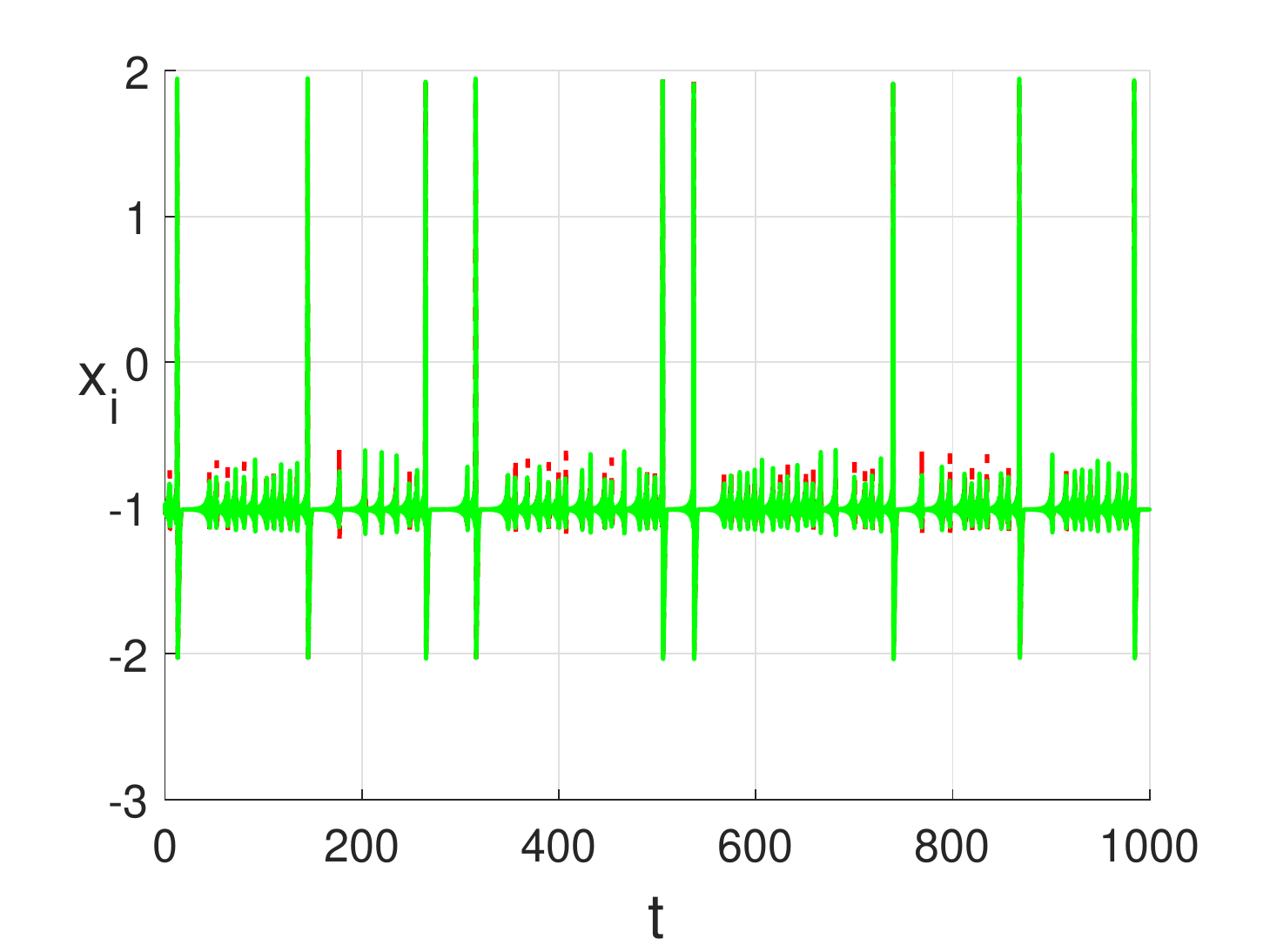}
    \label{TS_160_k1_0_094589_k2_45}
  }
  \caption{(a) Bifurcation tree. (b) Strange spiral attractor containing saddle-focus equilibrium with its homoclinic orbit, and (c) its enlarged part. (d) Time series for $k_2 = 45$ parameter. Parameters' values: $\alpha = 160^\circ$, $k_1=0.094589$.}
  \label{bif_k1_0_094589}
\end{figure}

Indeed, only for quite large values of parameter $k_2$ ($k_2 > 36$) regime $L_{hom}$ starts changing, see the bifurcation tree in figure~\ref{bif_tree_160_k1_0_094589}. Note that with further increase in $k_2$ interesting, similar to in-phase activity, chaotic regimes are developed from $L_{hom}$. This regime corresponds to the homoclinic spiral attractor, see figures~\ref{PP_160_k1_0_094589_k2_45} and \ref{PP_160_k1_0_094589_k2_45_part}. However the structure of this spiral attractor is more complicated comparing with the spiral attractor developed from $L_{12}$ (compare figures~\ref{PP_homoclinic_c} and \ref{PP_160_k1_0_094589_k2_45_part}). Time series corresponding to this spiral attractor (see figure~\ref{TS_160_k1_0_094589_k2_45}) confirms that such quasi-in-phase chaotic regime corresponds to the emergence of extreme event in the system. However, since such regimes are possible only for $k_2 > 41$, they are interesting only from mathematical point of view.

In the end of this section we note that parameter $C$ affects on all described regimes ($L_{anti}$, $L_{12}$, $L_{1221}$, $L_{hom}$) in the same manner as parameter $k_1$.

\section{Conclusion}
In this paper, we study the impact of electrical and memristor-based couplings on the regimes of in-phase, anti-phase and sequential activity, previously observed in the ensemble of two identical FitzHugh-Nagumo elements with chemical excitatory coupling \cite{KorotkovCNSNS2019}. We show that the presence of these couplings can lead to the appearance of new interesting regimes of neuron-like activity. In particular, we show that homoclinic spiral attractors corresponding to chaotic quasi-in-phase activity can appear in the system. We discuss that such regimes are dangerous for neurons since they can lead to the emergence of extreme events when interspike interval can suddenly become arbitrarily large.

However, in this paper we conduct only one-parameter bifurcation analysis. Comprehensive two-parameter bifurcation analysis is the subject of future work of the authors. These studies can be extremely useful for managing the complex collective behaviour of real neural systems.

Another promising direction for future research is related to the further development of our phenomenological model of minimal neuron ensemble. In particular, in our future work we will introduce to the model an external external periodic perturbation in order to simulate external control in the ensemble.

{\bf Acknowledgement.}

The authors thank S.V. Gonchenko for valuable advice and comments. This paper was supported by RSF grant 17-72-10228. Numerical experiments in Section 3 were supported by RFBR grant 18-31-20052. The authors also thank RFBR grants 19-01-00607, A. Kazakov thanks Basic Research Program at NRU HSE in 2019 for support of scientific researches.


\begin{thebibliography}{34}

\bibitem{Nicholls2011}
{J.G. Nicholls, A.R. Martin et al. From Neuron to brain. 5th ed. Sinauer Associates, (2011).}

\bibitem{pereda}			
{A.E.Pereda, Nature Reviews Neuroscience \textbf{15}, 250-263 (2014).}

\bibitem{sync}
{R. Reimbayev, I. Belykh, Int. J. Bifurc. Chaos \textbf{22}, 1440013 (2014).}

\bibitem{Afraimovich2004}
{V.S. Afraimovich, M.I. Rabinovich, P. Varona, Int. J. Bifurcation and Chaos \textbf{14(4)}, 1195-1208 (2004).}

\bibitem{Levanova2013}
{T.A. Levanova, M.A. Komarov, G.V. Osipov, Eur. Phys. J.: Special Topics \textbf{222(10)}, 2417-2428 (2013).}

\bibitem{chimera}
{B.K. Bera, D. Ghosh, M. Lakshmanan, Phys. Rev. E \textbf{93}, 012205 (2016).}

\bibitem{20}
{A. Riehle, S. Grun, M. Diesmann, Science \textbf{278}, 1950-1953 (1997).}

\bibitem{94}
{J. Fell, N. Axmacher, Nat. Rev. Neurosci. \textbf{12}, 105-118 (2011).}

\bibitem{21}
{S.M. Montgomery, G. Buzsaki, PNAS 104, 14495-14500 (2007).}

\bibitem{epilepsy}
{K. Lehnertz, S. Bialonski et al. Journal of Neuroscience Methods \textbf{183(1)}, 42-48 (2009).}

\bibitem{parkinson}
{I. Netoff, J. Schiff, J. Neurosci. \textbf{22}, 7297–7307 (2002).}

\bibitem{32}
{S. Majhi, M. Perc, D. Ghosh, Sci. Rep. \textbf{6}, 39033 (2016).}

\bibitem{33}
{Q.Y. Wang, M. Perc, et al. Phys. Rev. E \textbf{80}, 026206 (2009).}

\bibitem{34} 
{Q.Y. Wang, M. Aleksandra, M. Perc, Chinese Phys. B \textbf{20}, 040504 (2011).}

\bibitem{nosync}
{X.-J. Wang, J. Rinzel, Neural Comput. \textbf{4}, 84–97 (1002). }

\bibitem{151}
{B.W. Connors, M.A. Long, Annu. Rev. Neurosci. \textbf{27}, 393-418 (2004).}

\bibitem{223}
{H.V. Wheal, A.M. Thomson, Neuroscience \textbf{13}, 97-104 (1984).}

\bibitem{224}
{V. Zsiros, I. Aradi, G. Maccaferri, J. Physiol. \textbf{578}, 527-544 (2007).}

\bibitem{210}
{D.G. Placantonakis, A.A. Bukovsky, et al. J. Neurosci. \textbf{26}, 5008-5016 (2006).}

\bibitem{229}
{C.E. Landisman, B.W. Connors, Science \textbf{310}, 1809-1813 (2005).}

\bibitem{230}
{J.S. Haas, B. Zavala, C.E. Landisman, Science \textbf{334}, 389-393 (2011).}

\bibitem{231}
{J.S. Haas, C.M. Greenwald, A.E. Pereda, BMC Cell Biol. \textbf{17}, 51 (2016).}			

\bibitem{232}
{Z. Wang, R. Neely, C.E. Landisman, J. Neurosci. \textbf{35}, 7616-7625 (2015).}

\bibitem{233}
{J. O'Brien, Curr. Opin. Neurobiol. \textbf{29}, 64-72 (2014).}

\bibitem{187}
{S.A. Bloomfield, B. Volgyi, Nat. Rev. Neurosci. \textbf{10}, 495-506 (2009).}

\bibitem{234}
{B.W. Connors, Electrical signaling with neuronal gap junctions, in: A. Harris, D. Locke (Eds.), Connexins: A Guide, Humana Press/Springer 2009, pp. 143-164.}

\bibitem{235}
{A. Gelperin, J. Neurosci. \textbf{26}, 1663-1668 (2006).}

\bibitem{236}
{M.J. Kahana, J. Neurosci. \textbf{26}, 1669-1672 (2006).}

\bibitem{237}
{O. Paulsen, T.J. Sejnowski, J. Neurosci. \textbf{26}, 1661-1662 (2006).}

\bibitem{238}
{T.J. Sejnowski, O. Paulsen, J. Neurosci. \textbf{26}, 1673-1676 (2006).}

\bibitem{Belykh2017}
{R. Reimbayev, K. Daley, I. Belykh, Phil. Trans. R. Soc. A \textbf{375}, 20160282 (2017)}

\bibitem{239}
{N. Spruston, Neuron \textbf{31}, 669-671 (2001).}

\bibitem{240}
{W. Singer, C.M. Gray, Annu. Rev. Neurosci. \textbf{18}, 555-586 (1995).}

\bibitem{241}
{W. Singer, Neuron \textbf{24}, 49-65 (1999).}

\bibitem{MaTang2017}
{J.Ma, J.Tang, Nonlinear Dyn \textbf{89}, 1569-1578 (2017).}

\bibitem{LvWang2016}
{Lv, M., Wang, et al.  Nonlinear Dyn. \textbf{85}, 1479-1490 (2016).}

\bibitem{LvMa2016}
{Lv, M., Ma, J.	 Neurocomputing \textbf{205}, 375-381 (2016).}

\bibitem{Chua1971}
{Chua, L.O. IEEE Trans. Circuit Theory \textbf{18}, 507-519 (1971).}

\bibitem{13}
{L.O. Chua, Nanotechnology \textbf{24}, 383001/1-14 (2013).}

\bibitem{14}
{S.H. Jo, T. Chang, et al. Nano Letters \textbf{10}, 1297-1301 (2010).}

\bibitem{15}
{M. Laiho, E. Lehtonen, Proceedings IEEE of International Symposium on Circuits and Systems (ISCAS 2010), 2051-2054 (2010).}

\bibitem{16}
{B. Linares-Barranco, T. Serrano-Gotarredona, Proceedings of Nature, 1-4 (2009).}

\bibitem{37} 
{F.Q. Wu, C.N. Wang, et al. Physica A \textbf{469}, 81-88 (2017).}

\bibitem{Volos2015}
{Ch. K. Volos, I. M. Kyprianidis, et al., J. Engineering Science and Technology Review \textbf{8(2)}, 157-173 (2015).}

\bibitem{MaMi2017} 
{J. Ma, L. Mi, et al. Applied Mathematics and Computation \textbf{307}, 321-328 (2017).}

\bibitem{KorotkovCNSNS2019}
{A.G. Korotkov, A.O. Kazakov, et al. Commun.  Nonlin. Sci. Num. Simulat. \textbf{71}, 38-49 (2019).}

\bibitem{destexhe1994efficient}
{A. Destexhe, Z.F. Mainen, T.J. Sejnowski, Neural computation \textbf{6(1)}, 14-18 (1994).}

\bibitem{KorotkovIFAC2018}
{A.G. Korotkov, A.O. Kazakov, et al. IFAC-PapersOnLine \textbf{51(33)}, 241-245 (2018). }

\bibitem{Sharifi2010}
{M. J. Sharifi, Y. M. Banadaki, Journal of Circuits, Systems, and Computers \textbf{19(02)}, 407–424 (2010).}

\bibitem{FHN}
{Schwan H. P., ed.,	Biological engineering, McGraw-Hill Companies, 1969.}

\bibitem{Krupa2001}
{M. Krupa, P. Szmolyan, Journal of Differential Equations \textbf{174(2)}, 312–368 (2001).}

\bibitem{LevanovaPND}
{T.A. Levanova, A.O. Kazakov, et al. Izvestiya Vysshikh Uchebnykh Zavedeniy. Prikladnaya Nelineynaya Dinamika  \textbf{26(5)}, 101-112 (2018). }

\bibitem{Shil65}
{L.P. Shilnikov, Sov. Mat. Dok. \textbf{6}, 163 (1965).}

\bibitem{Shil70}
{L. P. Shilnikov, Mat. Sb. (N.S.) \textbf{81(123)},	92–103 (1970).}

\bibitem{extreme_event}
{A. Mishra, S. Saha, et al. Phys. Rev. E 97\textbf{}, 062311 (2018).}

\bibitem{GasGonTur}
{P. Gaspard, S. Gonchenko, Nonlinearity \textbf{10}, 409-423 (1997).}

\bibitem{long_interspike}
{D. S. Reich, F. Mechler, et al. Journal of Neuroscience \textbf{20(5)}, 1964-1974 (2000).}

\bibitem{BakKazKorLevOs18}
{Bakhanova, Y.V., Kazakov, et al. Eur. Phys. J. Special Topics \textbf{227(7-9)}, 959-970 (2018).}

\end{thebibliography}
\end{document}